\documentclass[preprint,12pt]{elsarticle}

\usepackage{amssymb}

\usepackage[modulo]{lineno}

\usepackage{amsmath}
\usepackage{graphicx}
\usepackage{subcaption} 
\usepackage{tikz}
\usepackage[font={small}]{caption}
\usepackage{comment}
\usepackage{siunitx}
\usepackage{enumitem}

\journal{Journal of Fluids and Structures}

\begin{document}

\begin{frontmatter}



\title{An investigation into the accuracy of the depth-averaging used in tidal turbine array optimisation}


\author[*]{Mohammad Amin Abolghasemi\corref{cor1}}
\ead{amin.abolghasemi@eng.ox.ac.uk}
\cortext[cor1]{Corresponding author}
\author[**]{Matthew D. Piggott}
\author[**]{Stephan C. Kramer}

\address[*]{Department of Engineering Science, University of Oxford, UK}
\address[**]{Applied Modelling and Computation Group, Department of Earth Science and Engineering, Imperial College London, UK}

\begin{abstract}
Depth-averaged shallow water models are widely used for the large-scale simulation of tidal turbine arrays. The relatively low computational complexity of this approach allows for layout optimisations aimed at improving the total array power output as well as an assessment of large-scale environmental impacts. In order to assess the suitability of using depth-averaged models to optimise array configurations, a comprehensive comparison between the wake profiles and power outputs predicted by a 2D shallow water model and a 3D actuator disc momentum (ADM) model is presented. Initially, a viscosity sensitivity analysis is presented to outline the limitations associated with using a constant eddy viscosity in the depth-averaged model and to outline the importance of correctly calibrating this value in line with the freestream velocity magnitude. Thereafter, the depth-averaged OpenTidalFarm (OTF) tool is used to optimise the positions of an array of 32 turbines in an ideal channel and the 3D Fluidity ADM-RANS model is used to assess the accuracy of the OTF predictions for the first time. It is shown that with the help of corrected power calculations a good agreement between the two models can be achieved, thus demonstrating the value of the eddy viscosity calibration implemented in the depth-averaged model.
\end{abstract}


\end{frontmatter}


\section{Introduction}
This study focuses on the extraction of tidal stream energy from coastal waters via horizontal axis tidal turbines which is currently one of the favoured approaches to efficiently harness the vast and reliably predictable tidal resource. The deployment of tidal turbines is a complex and expensive operation and this makes the task of locating the optimal positions for such turbines even more important.\\

Currently many large-scale marine hydrodynamic models employed to study marine energy use the depth-averaged shallow water equations, rather than the full 3D Navier-Stokes equations. Several site-specific numerical simulations have been documented. Sutherland et al. \cite{sutherland2007tidal} investigated the maximum tidal power potential of the Johnstone Strait, BC, Canada using a 2D finite element model (TIDE2D). Pham et al. \cite{pham2009tidal} demonstrated how the Telemac-2D shallow water solver could be used to perform a tidal resource assessment at the Paimpol-Br\'ehat site in France and Martin-Short et al. \cite{MartinShort2015596} used the Fluidity shallow water solver to model arrays of tidal turbines in the Inner Sound of the Pentland Firth. In order to numerically simulate arrays of turbines, these models usually adopt an approach where the turbines are represented as a region of increased bottom drag \cite{sutherland2007tidal, MartinShort2015596, Divett20120251}. Funke et al. exploited the relatively low complexity of such an approach to improve turbine positions to maximise array power generation using iterative adjoint-based optimisation techniques at a computational cost essentially independent of the number of turbines \cite{funke2014tidal}.\\

A main shortcoming of the approach is that it can fail to account for important turbulence physics and 3D effects, e.g. since the flow passing below and above the turbine is not modelled. 
In a realistic environment the ambient turbulence intensity has a significant effect on the structure of the turbine wake and its recovery \cite{blackmore2014influence, mycek2014experimental1, mycek2014experimental2}, which is of course crucial in an array design. However, when the depth-averaged shallow water equations are considered, the resulting turbine wake structures are strongly dependent on the viscosity coefficient used and this is often set to a spuriously high value in order to ensure a stable solution. The effect of ambient turbulence on the wake structure can therefore be misrepresented if an inappropriate viscosity value is used. Turbulence models for the shallow water equations have been suggested. For example, Nadaoka et al. \cite{nadaoka1998shallow} developed a model to simulate the evolution of horizontal large-scale eddies in shallow water by characterising the turbulence as a coexistence of 3D turbulence, with length scales smaller than water depth, and horizontal 2D eddies with much larger length scales. Moreover, Mungar \cite{MungarThesis2014} used a horizontal large eddy simulation (HLES) in combination with a 3D $k-\varepsilon$ model to examine the flow past tidal turbines using Delft3D \cite{hydraulics1999delft3d}. However, for the steady flow simulations presented therein the influence of HLES on the velocities was negligible. It was suggested that due to the implementation method of HLES in Delft3D, it is not effective for steady flows without velocity fluctuations and therefore the influence of HLES in combination with non-steady flows needs to be examined \cite{MungarThesis2014}.\\

Alternatively, the self-similar nature of far wake velocity profiles can be exploited to predict the flow past arrays of tidal turbines. Recently, Stallard et al. \cite{stallard2015experimental} investigated the mean wake properties behind a single three-bladed scaled turbine and demonstrated that for distances greater than 8 diameters downstream the velocity deficit becomes two-dimensional and self-similar. Stansby et al. \cite{stansby2016fast} extended this to demonstrate that the superposition of self-similar velocity profiles can lead to accurate predictions of depth-averaged wake velocities downstream of arrays of tidal turbines. This allowed for computationally efficient optimisation of turbine positions for power generation. However, although the method was shown to be reliable for up to three rows, it may not capture large scale wake behaviour which would be generated by multiple rows \cite{stansby2016fast}.\\

Another alternative approach would be to use a higher-fidelity 3D model coupled with an appropriate 3D turbulence model. A number of these models have been developed and validated against experimental flume tests with promising results \cite{roc2013methodology, afgan2013turbulent}, but their high computational expense has generally prevented their application in large-scale regional simulations and within iterative design optimisation. However, mesh optimisation techniques have the potential to help bridge the gap and improve the accuracy of large-scale simulations without the need for excessive computational power \cite{abolghasemi}.\\

In addition to this 2D vs 3D issue, Kramer et al. \cite{Kramer2016} recently examined the depth-averaged approach and pointed out that the resulting force exerted on the flow by a parameterised turbine agrees well with the theoretical value for coarse mesh resolutions only. As the mesh size becomes smaller than the length scale of the wake recovery, the exerted force starts decreasing with decreasing mesh sizes. The reason for this lies in the fact that the assumption that the upstream velocity can be approximated by the local model velocity, is no longer valid. In order to resolve this issue, Kramer et al. suggested using actuator disc momentum (ADM) theory to derive a correction to the enhanced bottom drag formulation. This leads to an improved estimate of the usefully extractable energy.\\

In order to assess the suitability of using depth-averaged models to optimise turbine array configurations, in this paper a depth-averaged model is used alongside a 3D hydrodynamic model based upon a Reynolds-averaged Navier-Stokes (RANS) approach with \textit{resolved} turbines using ADM theory. The sensitivity to the viscosity parameter employed in the depth-averaged model is investigated and consequently, with the aid of the ADM-RANS model, the viscosity value is tuned to improve the wake structure predicted by the lower fidelity model. Furthermore, the modifications suggested by Kramer et al. \cite{Kramer2016} are used to improve the thrust and power predictions of the depth-averaged solution. The combination of improved wake characterisation and more accurate thrust and power calculations in the depth-averaged model are then used to optimise turbine positions in order to maximise the total extracted power from an array. This is performed within the OpenTidalFarm (OTF) framework, an open source software tool for simulating and optimising tidal turbine arrays developed by Funke et al. \cite{funke2014tidal}. The depth-averaged results are then compared against 3D simulations of arrays of tidal turbines using an ADM-RANS model with mesh optimisation capabilities \cite{abolghasemi}, thus allowing for an investigation into the ultimate value of the depth-averaged adjoint-based optimisation used in OTF.\\

The paper is organised as follows. First in section \ref{sec:method} the depth-averaged formulation used in OTF and the 3D Fluidity ADM-RANS model with mesh optimisation capabilities are introduced. Descriptions of the thrust and power calculations in each model in line with the modifications suggested by Kramer et al. \cite{Kramer2016} are also presented. Then in section \ref{sec:nu_sens} the flow past a single turbine in an ideal channel is modelled using both models to determine their consistency and to arrive at a suitable viscosity value for use in OTF. This is then followed in section \ref{sec:channel} by an examination of the flow past an array of 32 tidal turbines in an ideal channel where the turbine positions are optimised for maximum power. The flow solution and the power gain predicted by OTF is compared against the values obtained from the 3D Fluidity ADM-RANS model in order to provide insight into the accuracy of the depth-averaged simulations. Initially steady flow scenarios are considered, before extending the analysis to examine an unsteady flow scenario with a time-dependent inlet velocity. The paper concludes with a general overview of the results.

  
\section{Methodology}\label{sec:method}

  \subsection{Depth-averaged model}
  OpenTidalFarm solves an optimisation problem constrained by the shallow water equations where the goal is to maximise power production $P$, i.e.

  \begin{align}
    \max_{\mathbf{m}}& \qquad P(\mathbf{m}, \mathbf{u}(\mathbf{m})) \, , \\
    \textnormal{subject to}& \qquad 
    \begin{cases}
      b_l \leq \mathbf{m} \leq b_u\\
      g(\mathbf{m}) \leq 0 \, ,
    \end{cases}
  \end{align}

  \noindent where $\mathbf{m}$ is a vector containing the turbine positions, $\mathbf{u}$ is the depth-averaged velocity, bounds $b_l \leq \mathbf{m} \leq b_u$ constrain the turbines to the (here rectangular) array area and the inequality constraint, $g(\mathbf{m}) \leq 0$, enforces a minimum distance spacing constraint between adjacent turbines. In each optimisation iteration, a two-dimensional finite element shallow water model predicts the hydrodynamics for given forcing and turbine locations, and thus the performance of the current array configuration can be evaluated by diagnosing the power produced. The gradient of the power extracted with respect to the turbine positions is then computed by solving the associated adjoint equations. These equations propagate causality backwards through the computation, from the power extracted back to the turbine positions. This yields the gradient at a cost almost independent of the number of turbines, which is crucial for any practical application targeted at large arrays \cite{funke2014tidal}. The optimisation is not limited to power production and can be used to maximise any functional of interest. More recently, Culley et al. \cite{Culley2016215} extended the model to include economic costs and hence to optimise the turbine positions to maximise profit over the lifespan of the array. In OTF the depth-averaged shallow water equations discretised are considered in the following form

  \begin{align}\label{eq:sw_eqs}
    \frac{\partial \mathbf{u}}{\partial t} + \mathbf{u}\cdot\nabla\mathbf{u} - \nu\nabla^2\mathbf{u} + g\nabla\eta + \frac{c_b+c_t(\mathbf{m})}{H}||\mathbf{u}||\mathbf{u} \, =& \, 0 \, , \\
    \frac{\partial \eta}{\partial t} + \nabla\cdot(H\mathbf{u}) \, =& \, 0 \, , \nonumber
  \end{align}

  \noindent where $\nu$ is the kinematic eddy viscosity, $\eta$ is the free surface displacement, $H$ is the total water depth, $g$ is the acceleration due to gravity, $c_b$ and $c_t(\mathbf{m})$ represent the background quadratic bottom friction and the local enhancement used to parameterise the  presence of turbines, respectively. A turbine is modelled via an increased bottom friction over a small area representative of an individual turbine. This is achieved via a \textit{bump} function which smoothly increases the friction value at the turbine locations:

  \begin{equation}\label{eq:bump_1d}
  \psi_{p,r} (x) = 
  \begin{cases}
    e^{1 - 1 / \left( 1 - ||\frac{x-p}{r}||^2 \right)} \qquad \textnormal{for}\, ||\frac{x-p}{r}||\,  <\,  1  \, ,\\
    0 \qquad \qquad  \qquad  \qquad \textnormal{otherwise} \, ,
  \end{cases}
  \end{equation}

  \noindent where $p$ and $r$ are the centre and the support radius of a 1D bump function, respectively. A two-dimensional bump function is obtained by multiplying Eq.~(\ref{eq:bump_1d}) by copies in both independent dimensions. The friction function of the $i$th turbine parameterised by friction coefficient $K_i$ centred at point $(x_i,y_i)$ is then given by

  \begin{equation}\label{eq:bump_2d}
    C_i(x,y) = K_i \, \psi_{x_i,r} (x) \, \psi_{y_i,r} (y) \, .
  \end{equation}

  \noindent The sum of the individual bottom friction fields associated with all $N$ turbines is denoted as  $c_t(\mathbf{m})$ in equation (\ref{eq:sw_eqs}) such that

  \begin{equation}\label{eq:ct}
    c_t(\mathbf{m}) = \sum_{i=1}^{N} C_i \, .
  \end{equation}

  \subsection{3D ADM-RANS model}
  For the purpose of the 3D simulations, an actuator disc model is used which utilises dynamic mesh optimisation techniques. The 3D model has been developed within the Fluidity framework, an open source finite element CFD code with 3D mesh optimisation capabilities \cite{piggott2008new}. The ADM-RANS model can therefore adapt the mesh dynamically in time and focus resolution only in the locations of interest. This allows for better management of limited computational resources without having to compromise on the accuracy of the solution, as demonstrated in \cite{abolghasemi}, making it particularly suitable for the study of large arrays of turbines. In the 3D ADM-RANS model, turbulence is accounted for by incorporating subgrid-scale models based on the Reynolds-averaged Navier-Stokes approach where the velocity is decomposed into mean, $\overline{\mathbf{u}}$, and fluctuating (turbulent), $\mathbf{u}'$, components, leading to the new momentum equation:
    
  \begin{equation}\label{eq:RANS}
    \frac{\partial \overline{\mathbf{u}}}{\partial t} + \overline{\mathbf{u}}\cdot\nabla\overline{\mathbf{u}} =
    -\frac{\nabla \overline{p}}{\rho} + \nu\nabla^2\overline{\mathbf{u}} - \nabla\cdot ( \overline{ \mathbf{u}' \otimes \mathbf{u}'} ) + S_u,
  \end{equation}

  \noindent where $\otimes$ denotes the outer product, $\overline{\mathbf{u}}$ is the mean velocity, $\mathbf{u}'$ is the fluctuating velocity, $\overline{p}$ is the mean pressure, $\rho$ is the fluid density and $S_u$ is the  momentum sink term included here to account for the presence of the turbines. The third term on the right hand side represents the effect of turbulent fluctuations on the mean flow and for incompressible flows is modelled as

  \begin{equation}\label{eq:R_tensor_incompressible}
    - \overline{ \mathbf{u}' \otimes \mathbf{u}'} = - \frac{2}{3} k\mathbf{I} + \nu_T \left( \nabla\overline{\mathbf{u}} + \left(\nabla\overline{\mathbf{u}}\right)^{\mathrm{T}} \right),
  \end{equation}

  \noindent where $k=(\overline{\mathbf{u}'\cdot\mathbf{u}'})/2$ is the turbulent kinetic energy and $\nu_T$ is the kinematic turbulent eddy viscosity. The momentum equations are closed by solving transport equations for $k$ and the turbulent frequency, $\omega$, which are used to obtain $\nu_T$, where the $k-\omega$ SST model has been used. For further details on the turbulence model used, the reader is referred to \cite{abolghasemi, menter2003ten}.\\

  The 3D numerical model incorporates turbines which are parameterised based on the ADM theory outlined by Houlsby et al. \cite{Houlsby2008}. ADM theory is based on the assumptions that the flow is inviscid and incompressible with uniform inflow. The turbine disc is infinitely thin and the thrust loading on the disc is uniformly spread. In the current model the circular disc has a small finite thickness and is represented as a piecewise constant scalar turbine field which is unity at the location of the disc and zero everywhere else in the domain. The ADM-RANS model uses a $\textnormal{P1}_{\textnormal{DG}}-\textnormal{P2}$ finite element pair \cite{cotter2009lbb} to discretise the RANS equations. This scheme uses the space of discontinuous piecewise linear functions ($\textnormal{P1}_{\textnormal{DG}}$) to represent velocity and the space of continuous piecewise quadratic functions ($\textnormal{P2}$) for pressure. A comparison against a finite volume OpenFOAM ADM model was presented in \cite{abolghasemi} to portray the benefits of the finite element discretisation scheme employed in the Fluidity ADM-RANS model. Therein, it was demonstrated that the Fluidity model performs better at capturing the sharp velocity variation across the actuator disc whereas the OpenFOAM model exhibits some fluctuations \cite{abolghasemi}.\\

  In order to set the appropriate loading on the disc, the Fluidity model uses the thrust coefficient, $C_t$, to compute the magnitude of thrust loading that should be applied at the disc. This is uniformly spread across the volume of the disc and is implemented as a momentum sink term in (\ref{eq:RANS}):

  \begin{equation}\label{eq:Su}
    S_u = - \frac{1}{V_t} \left( \frac{1}{2} \, A_t \, C_t \, u_0^2 \right) ,
  \end{equation}

  \noindent where $A_t$ is the cross-sectional area of the actuator disc, $V_t$ is the volume of the actuator disc over which the sink term is applied and $u_0$ is the unperturbed upstream streamwise component of velocity. In the context of the numerical simulations, it is important to ensure that $u_0$ is predicted accurately and in the 3D model it is computed using

  \begin{equation}\label{eq:u_t}
    u_t = \frac{1}{2} \left( 1 + \sqrt{1-C_t} \right) u_0 \, ,
  \end{equation}

  \noindent where $u_t$ is the average streamwise component of velocity over the elements making up the actuator disc \cite{hansen2000aerodynamics}.\\

  The Fluidity ADM-RANS model is capable of dynamic mesh optimisation which can be used to help reduce discretisation errors by refining the mesh in locations of numerical complexity or specific interest, e.g. regions with high velocity shear. In the interest of brevity, a detailed description of the mesh optimisation techniques is omitted here, and reference is made to Piggott et al. \cite{piggott2008new, Piggott2009} and Pain et al. \cite{pain2001tetrahedral}. For the purpose of this work the mesh is refined in regions of high curvature in the velocity, $k$ and $\omega$ fields, motivated by the desire to correctly capture the re-energisation of the wake downstream of the turbines \cite{abolghasemi}.

  \subsection{Thrust and power calculations}
  In order to ensure that the depth-averaged and fully 3D models are comparable, it is vital to ensure that firstly the same thrust is being applied in the different methods used to parameterise the presence of turbines, and secondly that the power calculation is consistent in both models. In the ADM-RANS model this is rather straightforward since the thrust applied is simply computed using

  \begin{equation}\label{eq:T3d}
    T_{3d} = \frac{1}{2}\rho\, A_t\, C_t\, u_0^2 ,
  \end{equation}

  \noindent and by assuming that the turbine power is equal to the product of the applied thrust and the local velocity, the extracted power is determined to be

  \begin{equation}\label{eq:P3d}
    P_{3d} = T_{3d} \times u_t.
  \end{equation}

  \noindent On the other hand, in OTF, the thrust applied is controlled by adjusting the amplitude of the bump function since

  \begin{equation}\label{eq:T2d}
    T_{2d} = \int_{A_c} \rho\, c_t\, u_c^2 ,
  \end{equation}

  \noindent where $c_t$ is the local friction coefficient enhancement used to parameterise the presence of turbines (\ref{eq:ct}), $u_c$ is the local depth-averaged velocity and $A_c$ is the area enclosed by the bump function. A corresponding formula for the extracted power in OTF, $P_{2d}$, is introduced later. Moreover, since (\ref{eq:T3d}) is expressed in terms of the unperturbed upstream velocity, $u_0$, in order to ensure consistency, a relationship between $u_c$ and $u_0$ is required. This issue has been recently discussed by Kramer et al. \cite{Kramer2016}, a brief overview of which is provided here.\\

  One of the main differences between the turbine representations in the ADM-RANS model and the depth-averaged model is that in the latter the turbine effectively blocks the entire channel depth. However, in the 3D ADM-RANS model, the horizontal and vertical components of velocity are resolved and the bypass flow passing underneath and above the turbine is also simulated. Hence, an equivalent thrust coefficient, $\hat{C_t}$, is defined for the depth-averaged model where\\

  \begin{equation}\label{eq:CT_hat}
    \hat{C_t} = \frac{A_t}{\hat{A_t}} \, C_t \, ,
  \end{equation}

  \noindent with

  \begin{equation*}
    \hat{A_t} = h \times w \, ,
  \end{equation*}

  \noindent where $h$ is the depth of the channel at the turbine location and $w$ is the width of the bump function. Hence, analogous to (\ref{eq:u_t}), the following relationship can be used to express the depth-averaged $u_c$ in terms of $u_0$:

  \begin{equation}\label{eq:u_c}
    u_c = \frac{1}{2} \left( 1 + \sqrt{1-\hat{C_t}} \right) \, u_0 \, .
  \end{equation}

  \noindent Note that this is the same $u_0$ used in (\ref{eq:Su}) and here the unperturbed 3D flow is assumed to be depth independent. Finally, equating the applied thrusts (\ref{eq:T2d}) to (\ref{eq:T3d}) and using (\ref{eq:u_c}) leads to

  \begin{equation}\label{eq:c_t}
    \int_{A_c} c_t = \frac{1}{2} A_t C_t \frac{4}{\left( 1 + \sqrt{1 - \frac{A_t}{\hat{A_t}} C_t} \right)^2} \, .
  \end{equation}

  \noindent Therefore, by using (\ref{eq:c_t}) $c_t$ can be set such that it yields a thrust in the depth-averaged model that matches the thrust applied in the ADM-RANS model. Furthermore, similar to the ADM-RANS model, the extracted power is the product of the thrust and the velocity at the location of the turbine. Hence, the depth-averaged velocity $u_c$ needs to be converted to an equivalent local turbine velocity analogous to $u_t$ used in (\ref{eq:P3d}). This can be achieved by combining (\ref{eq:u_t}) and (\ref{eq:u_c}) to yield

  \begin{equation}\label{eq:u_t2}
    u_t = \frac{1 + \sqrt{1 - C_t}}{1 + \sqrt{1 - \hat{C_t}}} \, u_c \, ,
  \end{equation}

  \noindent and this can be used to compute the extracted power using

  \begin{equation}\label{eq:P2d}
    P_{2d} = T_{2d} \times u_t \, .
  \end{equation}

\section{Viscosity sensitivity}\label{sec:nu_sens}
The ability to correctly account for the wake behind each turbine is of utmost importance when modelling arrays of tidal turbines. This was the main motivation behind the RANS approach adopted in the 3D Fluidity ADM-RANS model as the $k-\omega$ SST model used helps determine the wake length and momentum recovery depending on the ambient turbulence conditions \cite{abolghasemi}. On the other hand, in OTF the wake length can at present only be controlled via an eddy viscosity coefficient which in the current version of the software takes the same value everywhere in the domain. The reason for this lies in the fact that in order to keep iterative based turbine optimisation feasible, the computational cost behind each flow solve must be kept to a minimum and this limitation has not allowed for the inclusion of turbulence models in OTF thus far, although this is work in progress. However, by careful calibration of the viscosity coefficient, the disadvantages of not modelling for turbulence directly can be better understood and therefore minimised.\\

\begin{figure}[h!]
  \centering
  \begin{subfigure}[h!]{0.3\textwidth}
    \centering
    \includegraphics[scale=1]{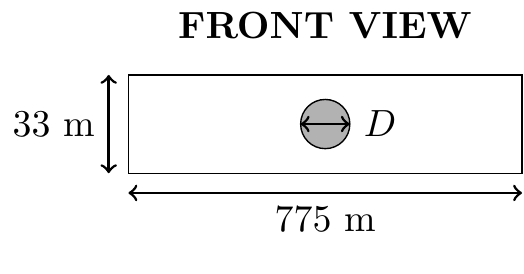}
  \end{subfigure} \hfill
  \begin{subfigure}[h!]{0.6\textwidth}
    \centering
    \includegraphics[scale=1]{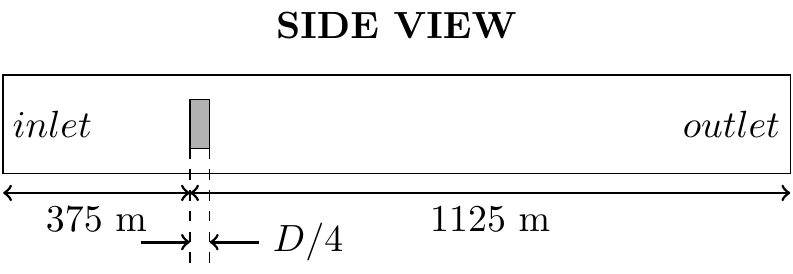}
  \end{subfigure}
  \caption{Numerical domain with turbine diameter $D=\SI{18.75}{\metre}$ and thickness $D/4$ used in the 3D ADM-RANS simulations. The dark grey area represents the actuator disc positioned at mid-depth. In the OTF simulations an equivalent 2D depth-averaged domain was used with a $D \times D$ square turbine drag area.}
  \label{fig:domain_single}
\end{figure}

In order to determine the significance of the viscosity coefficient in OTF a sensitivity study was carried out. Prior to proceeding with tidal turbine array simulations, the flow past a single turbine in an idealised channel is modelled using both the 3D Fluidity ADM-RANS model and the depth-averaged OTF package. The domain considered for this scenario is illustrated in Fig.~\ref{fig:domain_single}.\\

  \subsection{Boundary conditions}\label{sec:nu_sens_bc}
  In the ADM-RANS model, at the inlet, a Dirichlet boundary condition with constant inlet values, $u_\textnormal{in}$, $k_\textnormal{in}$ and $\omega_\textnormal{in}$ is applied, and the side walls and top surface are set to free slip. Furthermore, a zero flux boundary condition has been applied at the walls for both $k$ and $\omega$ and a zero pressure outflow boundary condition has been applied at the outlet. In order to simulate the flow inside the channel, it is important to capture the vertical asymmetry caused by the slower moving fluid near the bed. Hence, in the velocity field a quadratic drag boundary condition, with a non-dimensional drag coefficient of $C_D=0.0025$, is applied to the bottom surface. For more detail on the quadratic drag boundary condition used in the ADM-RANS model, the reader is referred to \cite{abolghasemi, fluiditymanual}.\\

  Similarly, in OTF, a Dirichlet velocity boundary condition with a constant inlet value is applied with the side walls set to free slip. The free surface displacement is set to zero at the outlet and a quadratic bottom friction $c_b=0.0025$ is prescribed. This drag is effectively applied across the whole \textit{depth} in the depth-averaged model, whereas the quadratic drag specified in the 3D ADM-RANS model is only applied at the bottom surface.\\

\begin{figure}[h!]
  \centering
  \includegraphics[width=0.8\textwidth]{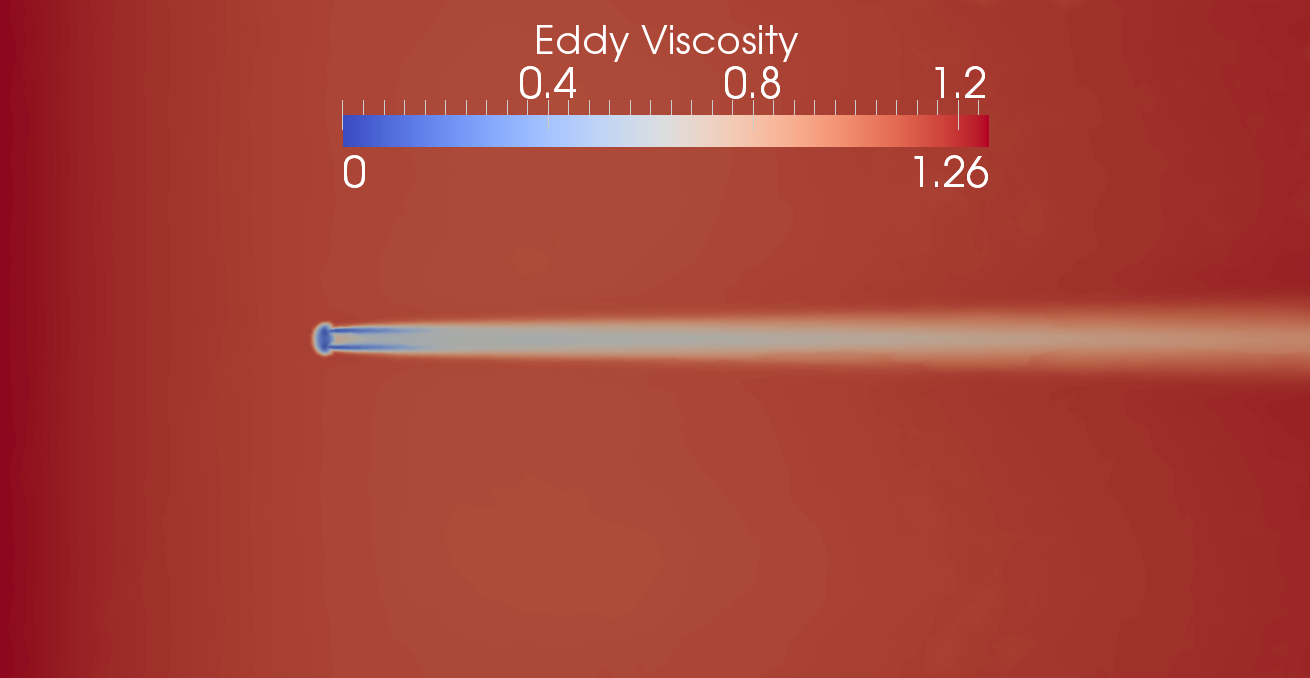}
  \caption{Variation of $\nu_T$ at mid-depth for the 3D ADM-RANS model with the domain used shown in Fig.~\ref{fig:domain_single}.}
  \label{fig:eddy}
\end{figure}

  Fig.~\ref{fig:eddy} illustrates the variation of $\nu_T$ at mid-depth in the 3D ADM-RANS model for the flow past a turbine with $u_\textnormal{in}=\SI{2.9}{\metre\per\second}$, $k_\textnormal{in}=\SI{0.126}{\square\metre\per\square\second}$, $\omega_\textnormal{in}=\SI{0.1}{\per\second}$ and $C_t=0.85$. Generally, the $\nu_T$ value is lower inside the wake downstream of the turbine. Furthermore, the smallest values are observed immediately upstream of the turbine and in the region between the slow moving fluid in the near wake and the accelerating bypass flow. These regions inherit the greatest levels of shear and therefore the ADM-RANS model has adjusted the $k$ and $\omega$ values such that $\nu_T$ is reduced at these locations. If the high $\nu_T$ present at the inlet was maintained everywhere, this would have encouraged more mixing and a shorter wake. Therefore, by reducing $\nu_T$ locally, the ADM-RANS model has delayed wake re-energisation. It was demonstrated in \cite{abolghasemi} that this approach can help predict correct wake lengths for different ambient turbulence values which agree with experimental observations. Note that this variation in eddy viscosity is not present in the depth-averaged OTF model since a constant value is used there.\\

  In order to accurately understand the consequences of using different viscosity coefficients, it was decided to compare the results from the Fluidity ADM-RANS model against those obtained via the depth-averaged OTF package for a range of different inlet conditions. This is crucial given the periodic nature of tidal flows which are of interest here. Therefore, a series of simulations with varying inlet velocities were carried out ranging from \SI{0.9}{\metre\per\second} to \SI{3.9}{\metre\per\second} using both models.\\

  The effect of turbulence and how the turbulence intensity ($I$) might change with varying velocities is also critical and therefore three different scenarios have been considered in the ADM-RANS simulations:

  \begin{enumerate}[label=(\alph*)]
    \item as $u_\textnormal{in}$ increases, both $k_\textnormal{in}$ and $\omega_\textnormal{in}$ are unaffected and therefore $\nu_T$ is unaffected but $I$ decreases
    \item as $u_\textnormal{in}$ increases,  $k_\textnormal{in}$ also increases but $\omega_\textnormal{in}$ is unaffected and therefore $I$ is unaffected but $\nu_T$ increases
    \item as $u_\textnormal{in}$ increases, both $k_\textnormal{in}$ and $\omega_\textnormal{in}$ also increase and therefore both $I$ and $\nu_T$ are unaffected
  \end{enumerate}

  Initially the inlet values are set to $u_\textnormal{in}=\SI{2.9}{\metre\per\second}$, $k_\textnormal{in}=\SI{0.126}{\square\metre\per\square\second}$ and $\omega_\textnormal{in}=\SI{0.1}{\per\second}$. This corresponds to $I=10\%$ and a turbulence length scale ($l$) of \SI{39.4}{\metre}, which is slightly larger than the channel depth. Consequently, as $u_\textnormal{in}$ is varied, $k_\textnormal{in}$ and $\omega_\textnormal{in}$ are, if necessary, adjusted relative to these values and in-line with the scenarios described above. This leads to various inlet conditions for scenarios (a)--(c) and the resulting inlet values are shown in Table~\ref{tab:nu_sens_a}--\ref{tab:nu_sens_c}. The $\nu_T$, $I$ and $l$ presented have been computed using

  \begin{align}\label{eq:aux}
    I     &= \frac{\sqrt{\frac{2}{3} k_\textnormal{in}}}{u_\textnormal{in}}  \, , \\
    \nu_T &= \frac{k_\textnormal{in}}{\omega_\textnormal{in}}                \, , \\
    l     &= \frac{\sqrt{k_\textnormal{in}}}{\beta^* \omega_\textnormal{in}} \, ,
  \end{align}
  
  \noindent with non-dimensional coefficient $\beta^* = 0.09$ \cite{menter2003ten}.\\
  
  \begin{table}[h!]
    \small
    \setlength{\tabcolsep}{8pt}
    \renewcommand{\arraystretch}{1.5}
    \begin{center}
      \begin{tabular}{|c c c c c c|}\hline
         $u_\textnormal{in}\,(\si{\metre\per\second})$ & $k_\textnormal{in}\,(\si{\square\metre\per\square\second})$ & $\omega_\textnormal{in}\,(\si{\per\second})$ & $I$  & $\nu_T$\,(\si{\square\metre\per\second}) & $l$\,(\si{\metre}) \\ \hline
         0.9  &  0.126  &  0.1  &  \textbf{0.32}  &  \textbf{1.26}  &  39.4  \\ 
         1.9  &  0.126  &  0.1  &  \textbf{0.15}  &  \textbf{1.26}  &  39.4  \\ 
         2.9  &  0.126  &  0.1  &  \textbf{0.10}  &  \textbf{1.26}  &  39.4  \\ 
         3.9  &  0.126  &  0.1  &  \textbf{0.07}  &  \textbf{1.26}  &  39.4  \\ \hline
      \end{tabular}
    \end{center}
    \caption{Different inlet conditions for scenario (a) of the ADM-RANS simulations.}
    \label{tab:nu_sens_a}
  \end{table}
  
  \begin{table}[h!]
    \small
    \setlength{\tabcolsep}{8pt}
    \renewcommand{\arraystretch}{1.5}
    \begin{center}
      \begin{tabular}{|c c c c c c|}\hline
         $u_\textnormal{in}\,(\si{\metre\per\second})$ & $k_\textnormal{in}\,(\si{\square\metre\per\square\second})$ & $\omega_\textnormal{in}\,(\si{\per\second})$ & $I$  & $\nu_T$\,(\si{\square\metre\per\second}) & $l$\,(\si{\metre}) \\ \hline
         0.9  &  0.012  &  0.1  &  \textbf{0.10}  &  \textbf{0.12}  &  12.2  \\ 
         1.9  &  0.054  &  0.1  &  \textbf{0.10}  &  \textbf{0.54}  &  25.8  \\ 
         2.9  &  0.126  &  0.1  &  \textbf{0.10}  &  \textbf{1.26}  &  39.4  \\ 
         3.9  &  0.228  &  0.1  &  \textbf{0.10}  &  \textbf{2.28}  &  53.1  \\ \hline
      \end{tabular}
    \end{center}
    \caption{Different inlet conditions for scenario (b) of the ADM-RANS simulations.}
    \label{tab:nu_sens_b}
  \end{table}
  
  \begin{table}[h!]
    \small
    \setlength{\tabcolsep}{8pt}
    \renewcommand{\arraystretch}{1.5}
    \begin{center}
      \begin{tabular}{|c c c c c c|}\hline
         $u_\textnormal{in}\,(\si{\metre\per\second})$ & $k_\textnormal{in}\,(\si{\square\metre\per\square\second})$ & $\omega_\textnormal{in}\,(\si{\per\second})$ & $I$  & $\nu_T$\,(\si{\square\metre\per\second}) & $l$\,(\si{\metre}) \\ \hline
         0.9  &  0.012  &  0.0095  &  \textbf{0.10}  &  \textbf{1.26}  &  128.1  \\ 
         1.9  &  0.054  &  0.043   &  \textbf{0.10}  &  \textbf{1.26}  &   60.0  \\ 
         2.9  &  0.126  &  0.1     &  \textbf{0.10}  &  \textbf{1.26}  &   39.4  \\ 
         3.9  &  0.228  &  0.181   &  \textbf{0.10}  &  \textbf{1.26}  &   29.3  \\ \hline
      \end{tabular}
    \end{center}
    \caption{Different inlet conditions for scenario (c) of the ADM-RANS simulations.}
    \label{tab:nu_sens_c}
  \end{table}

  Out of the three scenarios considered in the ADM-RANS simulations, scenario (c) appears to be the most realistic one. The reason for this lies in the fact that changes in the velocity fields will naturally alter the horizontal and vertical shear profiles within the flow and in order to capture this behaviour both $k$ and $\omega$ fields will have to be modified. Furthermore, it has been shown that increasing ambient $I$ leads to faster wake re-energisation \cite{mycek2014experimental1, mycek2014experimental2} and the importance of viscosity on the rate of wake re-energisation has already been stressed. Therefore, in scenario (c), $k_\textnormal{in}$ and $\omega_\textnormal{in}$ both increase with increasing $u_\textnormal{in}$ in order to maintain the same $I$ and $\nu_T$ at the inlet. Hence, $\nu_T$ at the inlet only changes if $I$ at the inlet changes and this reflects the close relationship between the two. However, one of the issues with scenario (c) is that, as a result of fixing $\nu_T$ at the inlet, the inlet $l$ values drop with increasing $u_\textnormal{in}$ and this is not physical. Alternatively, a scenario can be envisaged where $I$ and $l$ remains constant with increasing $u_\textnormal{in}$, but in that case $\nu_T$ at the inlet will no longer remain constant. Generally, at a realistic tidal site the change in flow features will be more complicated than the constant turbulent intensity assumption used here and the variations in $k$ and $\omega$ will depend on the particular site under consideration.

  \subsection{Results}\label{sec:nu_sens_results}
  In the 3D ADM-RANS model a turbine with $C_t=0.85$ and $D=\SI{18.75}{\metre}$ has been assumed. In OTF the turbine cell area, $A_c$, where the bump function is defined is a square region with side $D$ explicitly resolved by multiple triangles in the unstructured mesh. In order to ensure the same thrust is applied in both models, Eq.~(\ref{eq:c_t}) is used to compute the equivalent OTF parameter to be $\int_{A_c} c_t=0.418$. Consequently, this requires $K=1.147$ in (\ref{eq:bump_2d}). In the OTF simulations a range of different eddy viscosity values ranging from \SI{0.1}{\square\metre\per\second} to \SI{1}{\square\metre\per\second} have been considered in increments of \SI{0.1}{\square\metre\per\second}. This will help outline the limitations of using the same viscosity value everywhere for the different scenarios considered. This range of eddy viscosity values has been chosen since it is similar to the range of $\nu_T$ values observed in the ADM-RANS simulations, Fig.~\ref{fig:eddy}.\\

  In the 3D ADM-RANS simulations, mesh optimisation has been used to refine the mesh in regions of high curvature in the velocity, $k$, and $\omega$ fields. The minimum and maximum values of the element edge length ($l_e$) were set to $l_e / D = 0.125$ and $l_e / D = 2$, respectively. The results from the 3D ADM-RANS simulations are then depth-averaged in a post-processing step to allow for a comparison against the OTF results. In OTF, an unstructured fixed 2D mesh has been used with minimum $l_e / D = 0.125$ near the turbine and maximum $l_e / D = 1.07$ at the boundaries. The minimum value used is the same as the one used in the ADM-RANS model and the maximum value is almost half the value used in the 3D model.\\

  \begin{figure}[h!]
    \centering
    \includegraphics[scale=0.62]{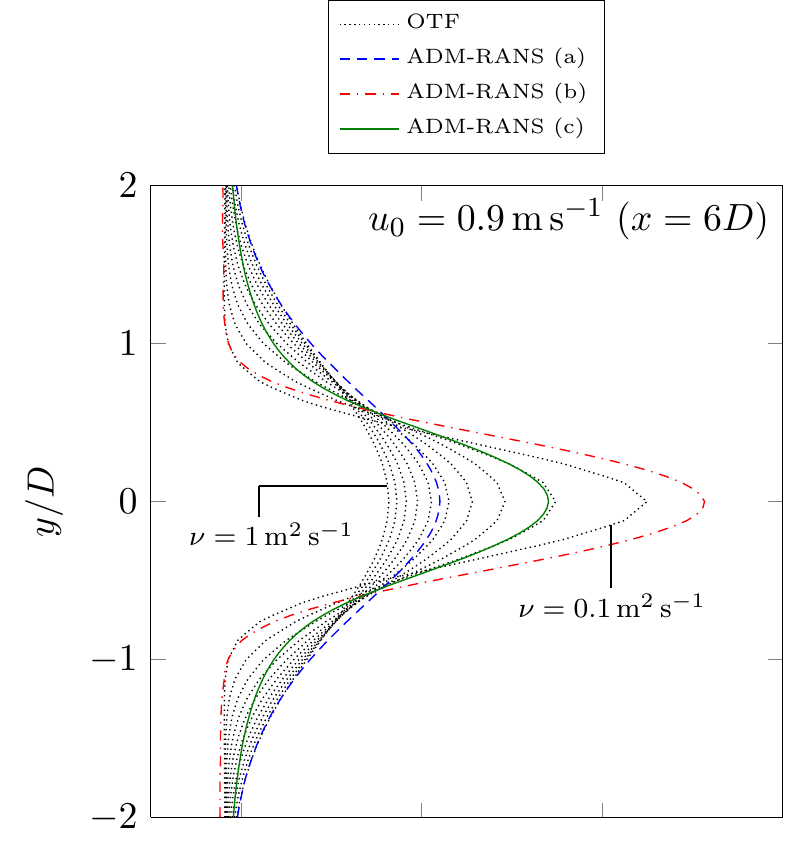}\hfil
    \includegraphics[scale=0.62]{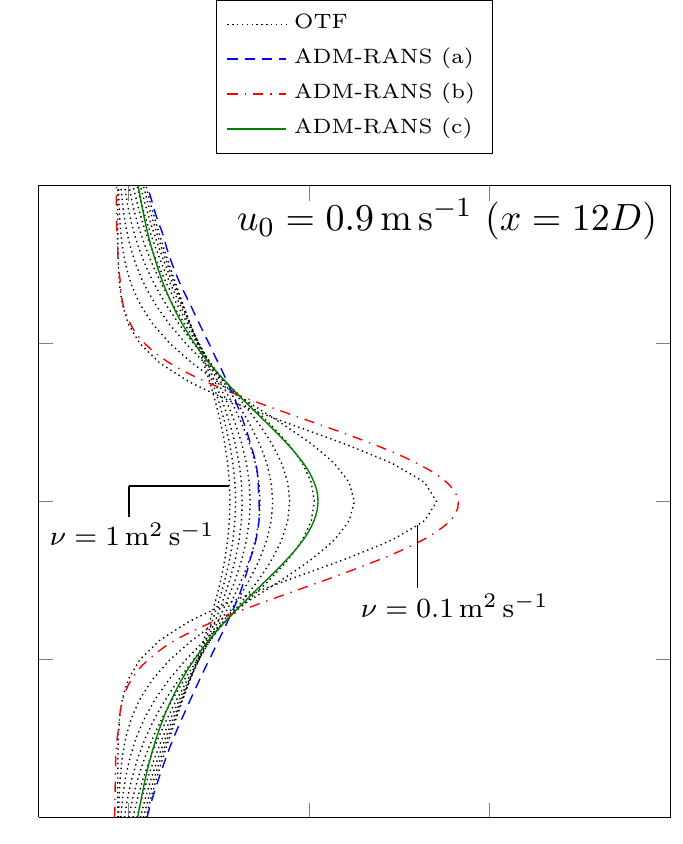}
    \includegraphics[scale=0.62]{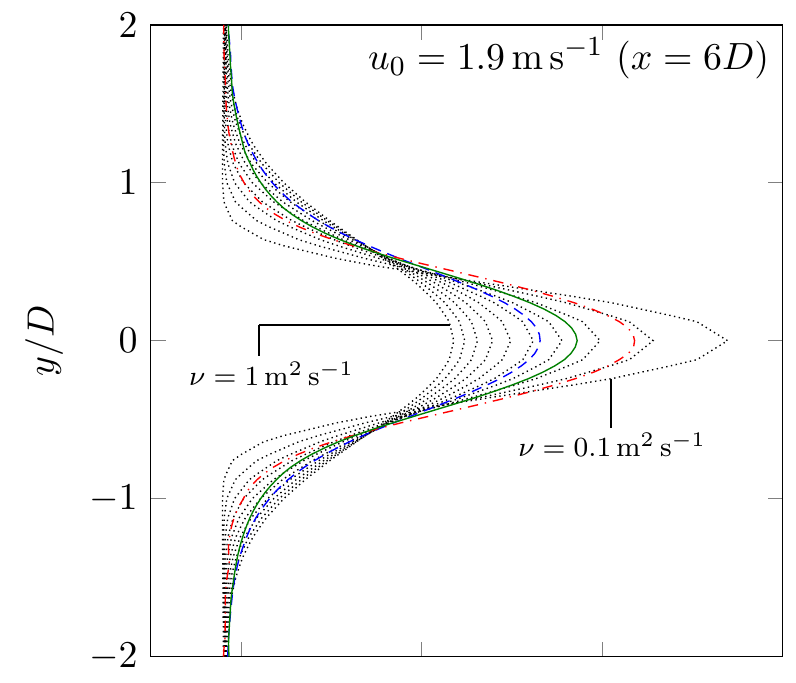}\hfil
    \includegraphics[scale=0.62]{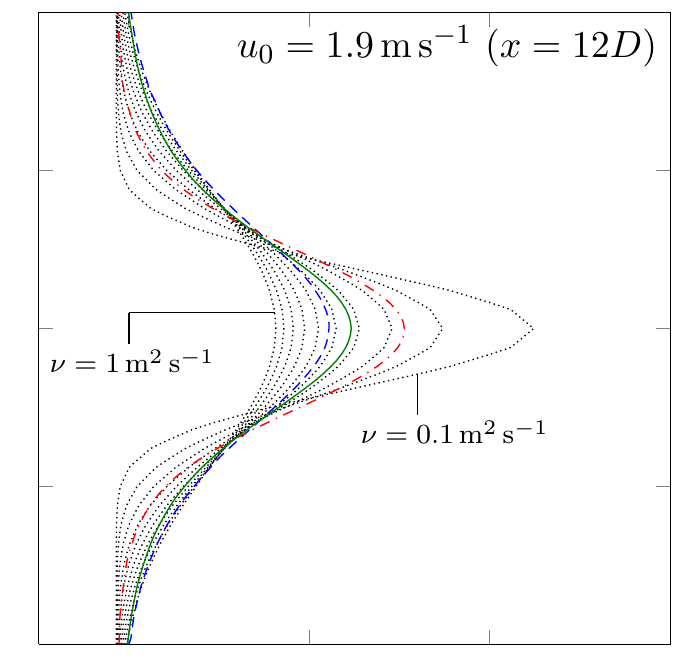}
    \includegraphics[scale=0.62]{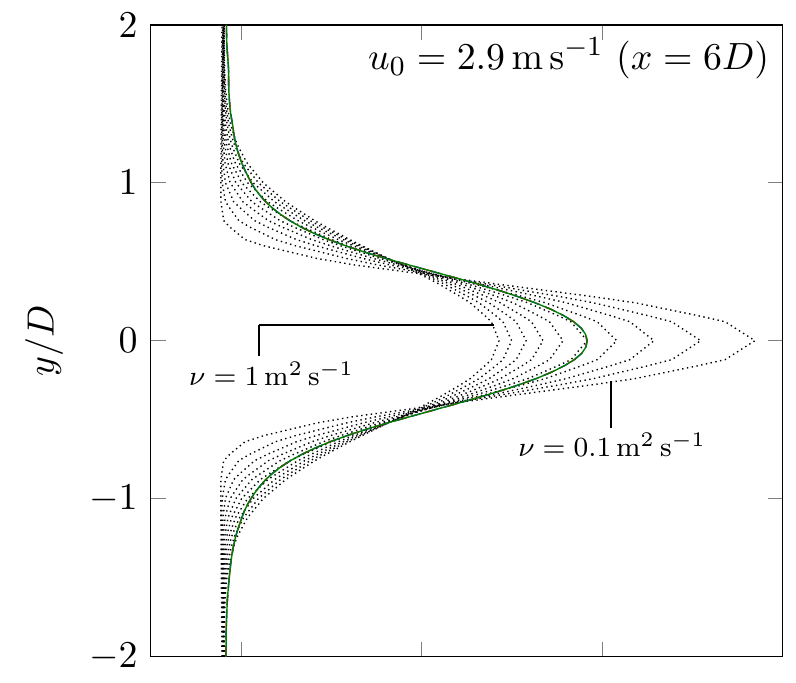}\hfil
    \includegraphics[scale=0.62]{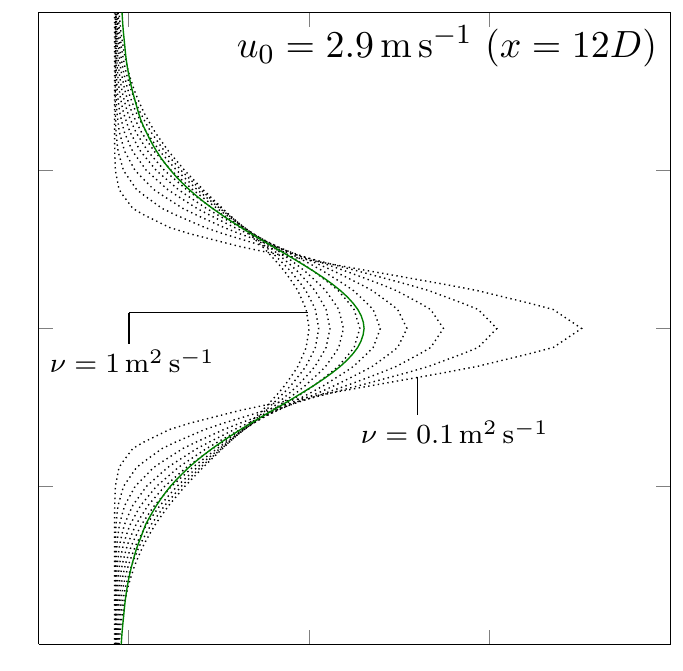}
    \includegraphics[scale=0.62]{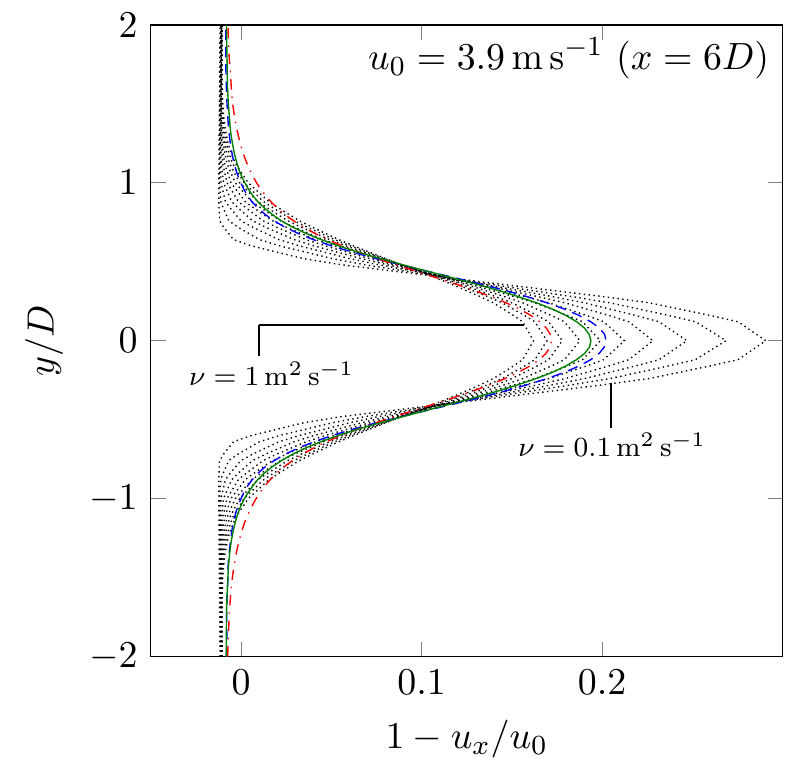}\hfil
    \includegraphics[scale=0.62]{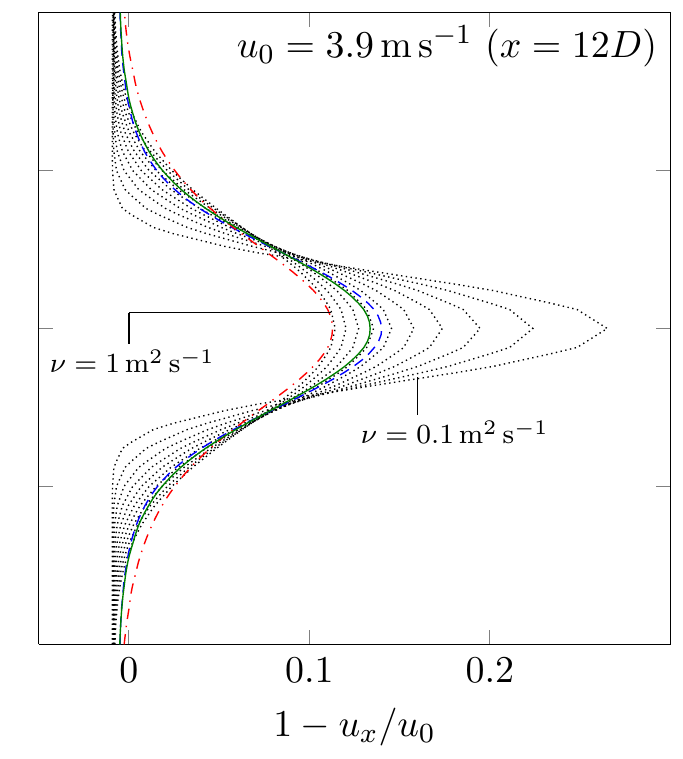}

    \caption{Lateral velocity deficit profiles at $6D$ and $12D$ at four different $u_\textnormal{in}$ values for the OTF and the Fluidity ADM-RANS simulations. The dotted lines represent the OTF results where $\nu$ is increased from \SI{0.1}{\square\metre\per\second} to \SI{1}{\square\metre\per\second}. The 3D ADM-RANS results have been depth-averaged and are plotted on top of the OTF results to help identify the OTF $\nu$ value that best matches the ADM-RANS profiles.}
    \label{fig:nu_sens}
  \end{figure}

  Fig.~\ref{fig:nu_sens} displays lateral velocity deficit profiles at $6D$ and $12D$ downstream of the turbine for the various OTF and ADM-RANS simulations at four different $u_\textnormal{in}$ values. The dotted lines represent the OTF results for different eddy viscosity values. As $\nu$ increases from \SI{0.1}{\square\metre\per\second} to \SI{1}{\square\metre\per\second}, the velocity deficit in the wake is reduced and the lateral expansion of the wake increases as expected. Furthermore, for the same $\nu$ value in OTF, the wake length grows as $u_\textnormal{in}$ increases from \SI{0.9}{\metre\per\second} to \SI{3.9}{\metre\per\second}. This is reflected in the higher velocity deficit values observed. Also note that the lateral width of the wake decreases with increasing $u_\textnormal{in}$.
This is due to the fact that as $u_\textnormal{in}$ increases the difference in velocity between the bypass flow and the wake also increases and this delays wake re-energisation.\\

  The results from the three scenarios considered using the ADM-RANS model have been depth-averaged and are plotted on top of the OTF results. This will help contrast the OTF results against those obtained using the 3D ADM-RANS model. The three different scenarios considered in the ADM simulations lead to significantly different results. In scenarios (a) and (c) the behaviour is similar to the OTF results in that the wake lengths grow with increasing $u_\textnormal{in}$. This growth is more pronounced in scenario (a) than in scenario (c), where the wake lengths are not significantly affected by the increase in $u_\textnormal{in}$. In scenario (b) however, the opposite can be observed where the deficit decreases with increasing $u_\textnormal{in}$. Scenario (b) is the only scenario where the inlet $\nu_T$ is changing with $u_\textnormal{in}$, in the other two scenarios $\nu_T$ at the inlet is fixed. Hence, compared to the other two scenarios, the inlet $\nu_T$ is lower at $u_\textnormal{in}=\SI{0.9}{\metre\per\second}$. This indicates less mixing between the bypass flow and the wake, which leads to longer wakes and greater velocity deficit values. As $u_\textnormal{in}$ increases to \SI{3.9}{\metre\per\second}, the inlet $\nu_T$ also increases and this encourages more mixing and shorter wake profiles. Furthermore, although the inlet $\nu_T$ value of \SI{1.26}{\square\metre\per\second} present in the 3D ADM-RANS simulations is outside the range of $\nu$ values considered in OTF, it has been shown that a value of \SI{1}{\square\metre\per\second} overpredicts the rate of wake re-energisation, Fig.~\ref{fig:nu_sens}, and increasing the OTF $\nu$ value further will only lead to a larger discrepancy between the two models.\\

\begin{figure}[h!]
    \centering
    \includegraphics[scale=1]{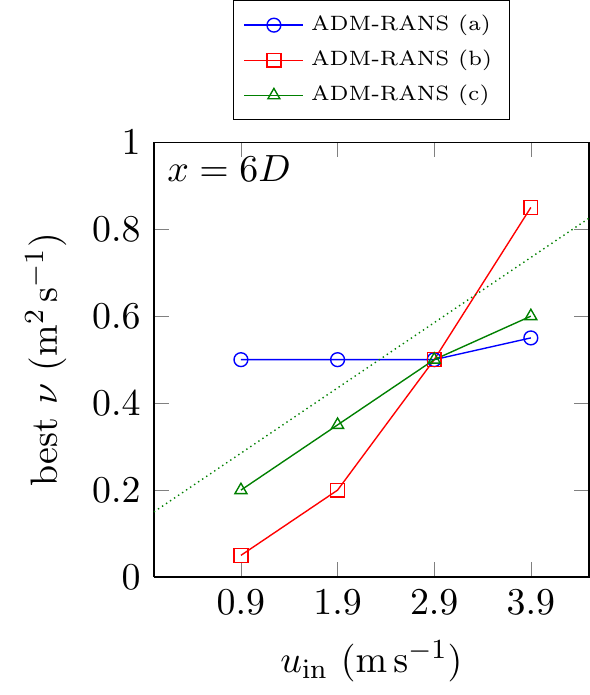}\hfil
    \includegraphics[scale=1]{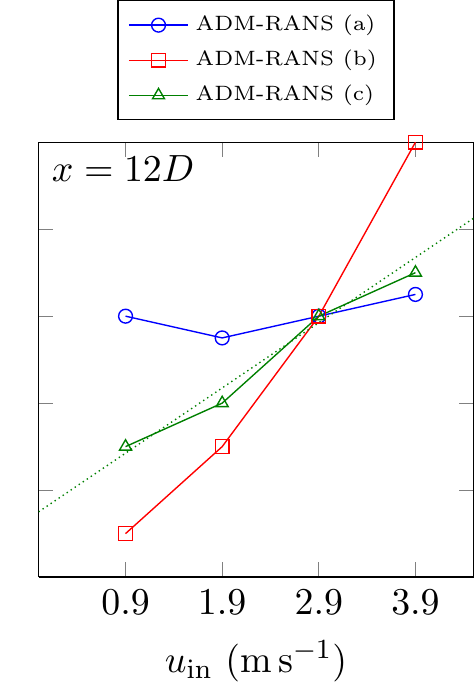}
    \caption{Based on the wake profiles presented in Fig.~\ref{fig:nu_sens}, the OTF $\nu$ value that provides the closest match to the ADM-RANS profile has been selected for each ADM-RANS simulation and the results are presented here. The dotted lines show the linear relationship described in Eq.~(\ref{eq:nu_best}).}
    \label{fig:best_nu}
  \end{figure}

  Overall, Fig.~\ref{fig:nu_sens} highlights the shortcomings that result due to the fixed viscosity constraint in the OTF simulations given that there is not an OTF setup (i.e. fixed $\nu$ value) that agrees with the ADM-RANS profiles perfectly. This suggests that even if a constant eddy viscosity is to be used everywhere in the domain, this value should be at least adjusted in-line with the upstream velocity value or the point in the tidal cycle. In order to shed light on the relationship between $u_\textnormal{in}$ and the most suitable viscosity value, the OTF viscosity value that leads to the best match velocity deficit profile for each ADM-RANS run has been selected and the results are presented in Fig.~\ref{fig:best_nu}.\\

  In order to come up with a simple relationship between $u_\textnormal{in}$ and the most appropriate $\nu$ (recalling that only \SI{0.1}{\square\metre\per\second} increments in values were considered) to be used in OTF, scenario (c) is the one considered as it is the most realistic scenario. Hence, the following linear relationship is suggested where

  \begin{equation}\label{eq:nu_best}
    \nu = \nu_\textnormal{min} + l_\nu \, u_\textnormal{in} \, ,
  \end{equation}

  \noindent with $l_\nu = \SI{0.15}{\metre}$ and $\nu_\textnormal{min} = \SI{0.15}{\square\metre\per\second}$ based on the results presented in Fig.~\ref{fig:best_nu}. This relationship provides a closer match at $x=12D$ than at $x=6D$ for the scenario (c) dataset. Given that this study focuses on array modelling, the far wake profiles are deemed of greater importance and the coefficients in Eq.~(\ref{eq:nu_best}) have been chosen with that in mind.

\section{Channel flow}\label{sec:channel}
Having established the means to address the key differences in the wake profiles predicted by the depth-averaged OTF model and the 3D ADM-RANS model, the flow past an array of 32 tidal turbines was considered using both models in order to assess the suitability of the adjoint-based optimisation used in OTF. Initially, steady flow cases are considered with constant inlet velocities, and then an unsteady case is presented with a time-dependent inlet velocity.

  \subsection{Steady flow}\label{sec:channel_steady}
  The same ideal channel of the previous section is used here with the same boundary conditions described in section \ref{sec:nu_sens_bc}. Two different inlet velocity values of $\SI{1.9}{\metre\per\second}$ and $\SI{2.9}{\metre\per\second}$ have been considered to allow for a comprehensive comparison between the two models. In this section, for the depth-averaged OTF simulations the steady state shallow water equations are solved and for the 3D ADM-RANS simulations the unsteady RANS equations are run until steady state is achieved.

  \subsubsection{OpenTidalFarm -- depth-averaged}
\begin{figure}[h!]
    \centering
    \includegraphics[scale=1]{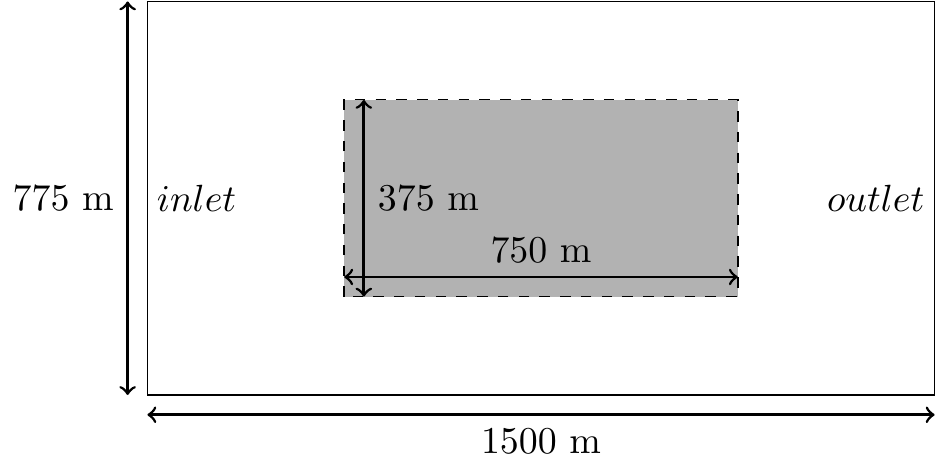}
    \caption{OTF domain for an array of 32 turbines in an ideal channel where the grey area represents the site where turbines can be positioned.}
    \label{fig:domain_32}
  \end{figure}

  Initially the 32 turbines were arranged in 9 rows in a staggered layout and OTF was used to maximise the extracted power from the array by optimising the turbine positions. The eddy viscosity coefficient used has been calculated using Eq.~(\ref{eq:nu_best}) and therefore $\nu=\SI{0.435}{\square\metre\per\second}$ and $\nu=\SI{0.585}{\square\metre\per\second}$ for $u_\textnormal{in}=\SI{1.9}{\metre\per\second}$ and $u_\textnormal{in}=\SI{2.9}{\metre\per\second}$, respectively. In these simulations the 32 turbines are assumed to be identical and the turbine properties (i.e. $A_c$ and $K$) are the same as those used in section \ref{sec:nu_sens_results}. The turbines were constrained to a $\SI{750}{\metre}\times\SI{375}{\metre}$ rectangular area in the middle of the channel as illustrated in Fig.~\ref{fig:domain_32}. The OTF package has been used to optimise the position of the turbines within this area in order to maximise the power output of the array of 32 turbines in each case. A minimum spacing of $2D$ between adjacent turbines was enforced in the OTF optimisation.\\

\begin{figure}[h!]
    \centering
    \begin{subfigure}[h!]{0.45\textwidth}
      \centering
      \includegraphics[width=\textwidth]{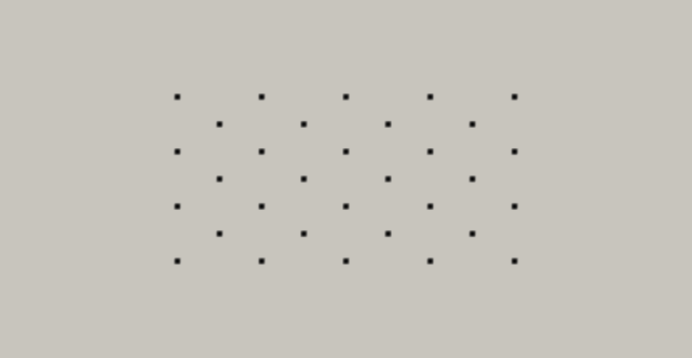}
      \caption{staggered turbine positions}
    \end{subfigure} \hfill
    \begin{subfigure}[h!]{0.45\textwidth}
      \centering
      \includegraphics[width=\textwidth]{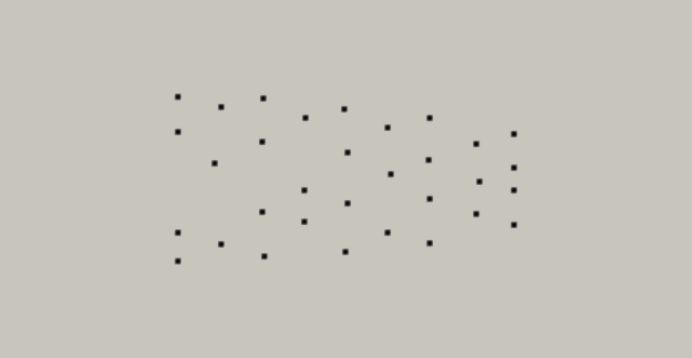}
      \caption{optimised turbine positions}
    \end{subfigure}

    \begin{subfigure}[h!]{0.45\textwidth}
      \centering
      \includegraphics[width=\textwidth]{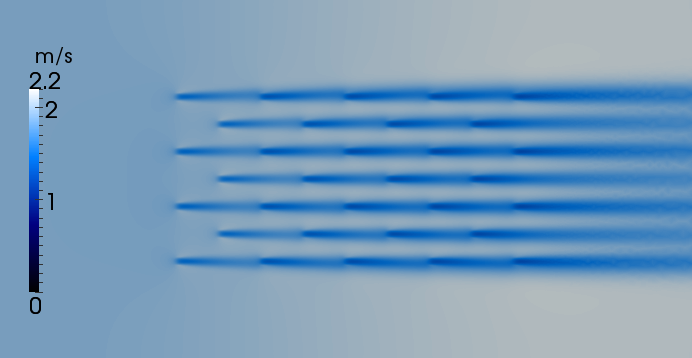}
      \caption{staggered velocity magnitude}
    \end{subfigure} \hfill
    \begin{subfigure}[h!]{0.45\textwidth}
      \centering
      \includegraphics[width=\textwidth]{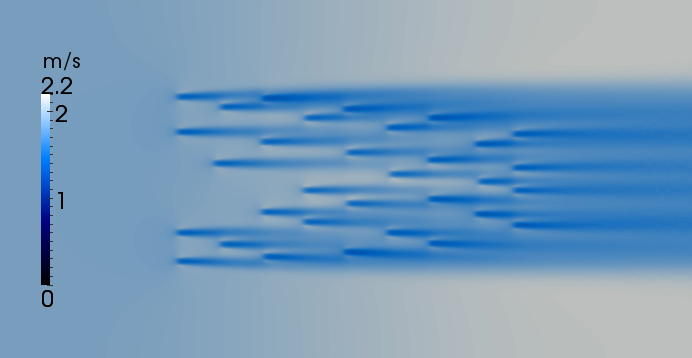}
      \caption{optimised velocity magnitude}
    \end{subfigure}
    \caption{OTF results showing the initial staggered layout (\SI{12.98}{\mega\watt}) on the left and the optimised layout (\SI{17.12}{\mega\watt}) on the right for the $u_\textnormal{in}=\SI{1.9}{\metre\per\second}$ case.}
    \label{fig:OTF_32_pos_19}
  \end{figure}

\begin{figure}[h!]
    \centering
    \begin{subfigure}[h!]{0.45\textwidth}
      \centering
      \includegraphics[width=\textwidth]{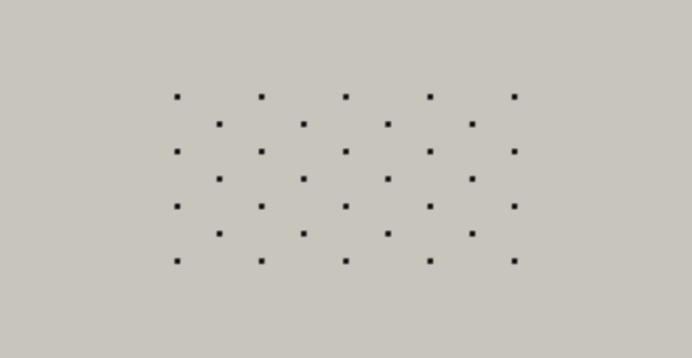}
      \caption{staggered turbine positions}
    \end{subfigure} \hfill
    \begin{subfigure}[h!]{0.45\textwidth}
      \centering
      \includegraphics[width=\textwidth]{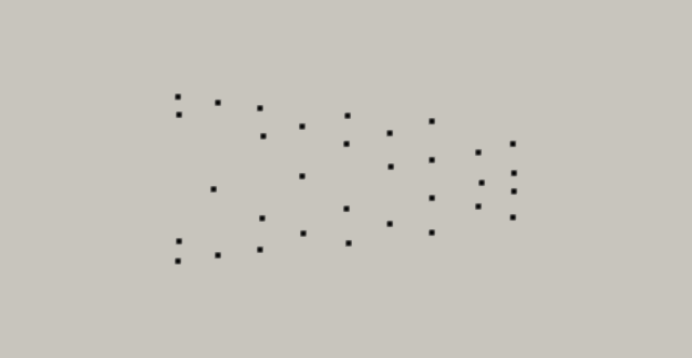}
      \caption{optimised turbine positions}
    \end{subfigure}

    \begin{subfigure}[h!]{0.45\textwidth}
      \centering
      \includegraphics[width=\textwidth]{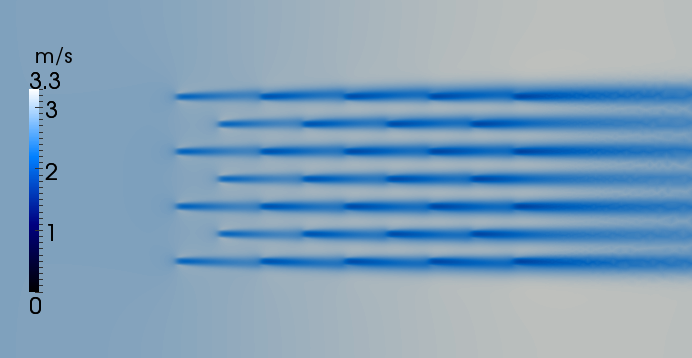}
      \caption{staggered velocity magnitude}
    \end{subfigure} \hfill
    \begin{subfigure}[h!]{0.45\textwidth}
      \centering
      \includegraphics[width=\textwidth]{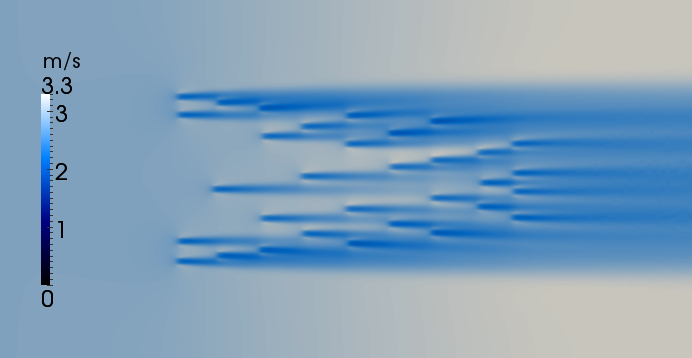}
      \caption{optimised velocity magnitude}
    \end{subfigure}
    \caption{OTF results showing the initial staggered layout (\SI{45.48}{\mega\watt}) on the left and the optimised layout (\SI{61.93}{\mega\watt}) on the right for the $u_\textnormal{in}=\SI{2.9}{\metre\per\second}$ case.}
    \label{fig:OTF_32_pos_29}
  \end{figure}

  The optimisation results are shown in Fig.~\ref{fig:OTF_32_pos_19}--\ref{fig:OTF_32_pos_29} with the initial staggered turbine layout on the left and the optimised turbine layout on the right. The total array extracted power has been computed using Eq.~(\ref{eq:P2d}) and the results are shown in Table~\ref{tab:power_comp}. In both cases the optimised layout leads to substantial increases in total power with a 32\% increase for the $u_\textnormal{in}=\SI{1.9}{\metre\per\second}$ case and a 37\% increase for the $u_\textnormal{in}=\SI{2.9}{\metre\per\second}$ case. The two optimised layouts are very similar and in both cases OTF has moved the turbines, mainly in the lateral direction, in order to make sure no turbine is placed in the wake of an upstream turbine. This ensures a higher upstream velocity, which will in turn result in greater extracted power by the turbine.\\

  \subsubsection{Fluidity -- 3D ADM-RANS}
  The Fluidity ADM-RANS model has been used to check the power predictions of OTF and the suitability of the eddy viscosity values used in the simulations of the array of 32 turbines. Hence, the flow past the initial staggered layouts, the final optimised layouts as well as two randomly chosen intermediate layouts, have been simulated using the 3D ADM-RANS model. The inlet turbulence properties have been set using scenario (c), described previously in section \ref{sec:nu_sens_bc}. As with the OTF simulations, the 32 turbines are assumed to be identical with $C_t=0.85$. Note that $C_t$ is assumed to be constant in this work; however \textit{thrust curves}, where $C_t$ is expressed as a function of $u_0$, could also be easily used instead. This represents the performance of a real device more accurately as it enables the model to take into account the cut-in speed below which the turbine does not operate and the rated speed above which $C_t$ decreases to maintain a constant power yield \cite{MartinShort2015596}. This would be a worthy addition to the model and would be necessary when simulating realistic scenarios.\\

  The mesh optimisation capabilities of the Fluidity ADM-RANS model makes these simulations feasible since there is no need to produce a fine mesh prior to the simulations, anticipating the location of the wakes (which in a later solution vary both in space as well as time). This is especially true when considering the optimised layouts, where the turbine layout is irregular. Once again the mesh is refined in regions of high curvature in the velocity, $k$, and $\omega$ fields and the edge lengths values used were identical to those described in section \ref{sec:nu_sens_results}.\\

\begin{figure}[h!]
    \centering
    \begin{subfigure}[h!]{0.8\textwidth}
      \centering
      \includegraphics[width=\textwidth]{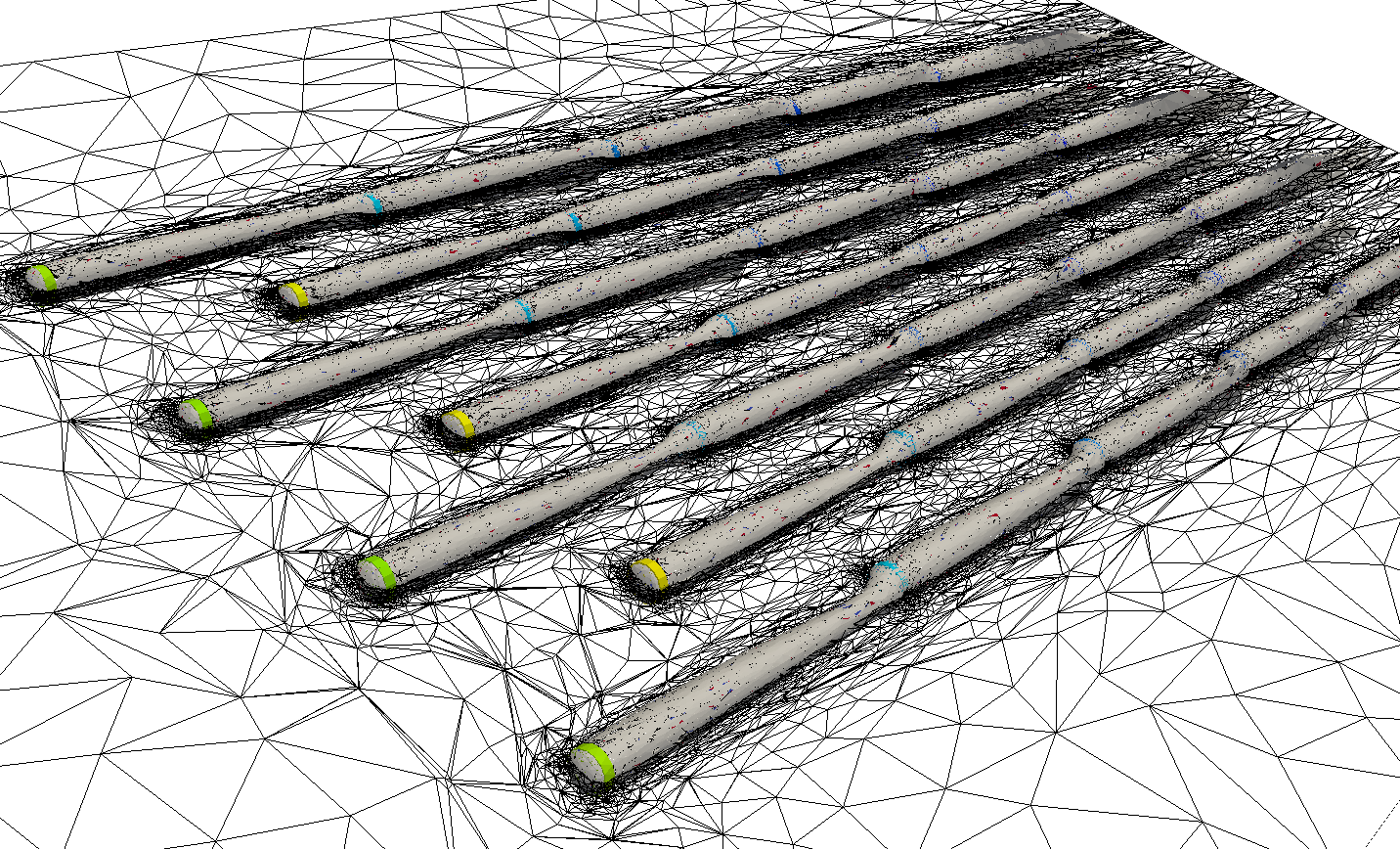}
      \caption{staggered layout (\SI{46.48}{\mega\watt})}
    \end{subfigure} \hfill
    \begin{subfigure}[h!]{0.8\textwidth}
      \centering
      \includegraphics[width=\textwidth]{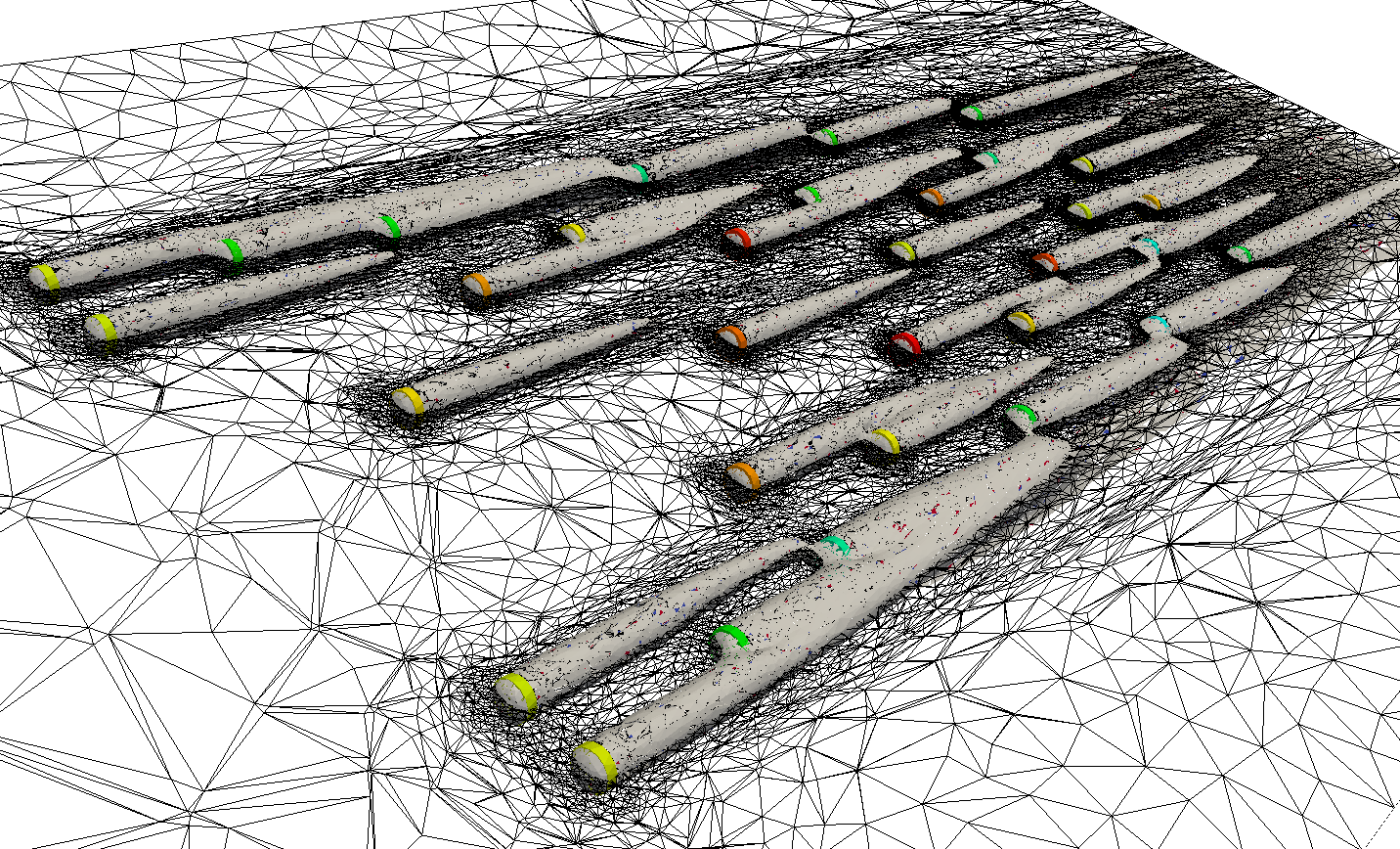} 
      \caption{optimised layout (\SI{59.12}{\mega\watt})}
    \end{subfigure}
    \caption{Iso-surfaces of constant speed, $||\textbf{u}||=0.85 \, u_\textnormal{in}$, are shown here to illustrate the wakes formed behind the turbine array in both the staggered and optimised layouts for the Fluidity ADM-RANS simulations with $u_\textnormal{in}=\SI{2.9}{\metre\per\second}$. A 2D slice at hub height through the 3D domain is also shown to demonstrate the optimised mesh where a higher resolution is used near the turbines and their wakes. The turbine colours correspond to the relative power output with red being the highest and blue corresponding to the lowest.}
    \label{fig:FL_32}
  \end{figure}

\begin{figure}[h!]
    \centering
    \includegraphics[scale=0.8]{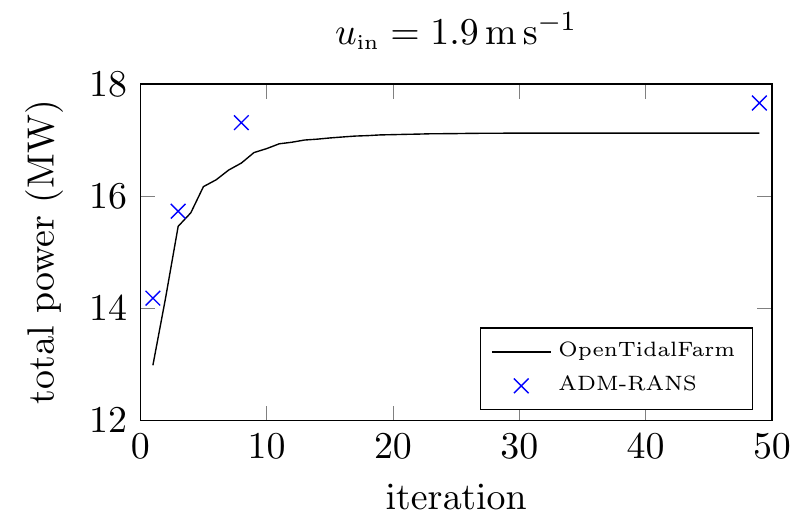}
    \includegraphics[scale=0.8]{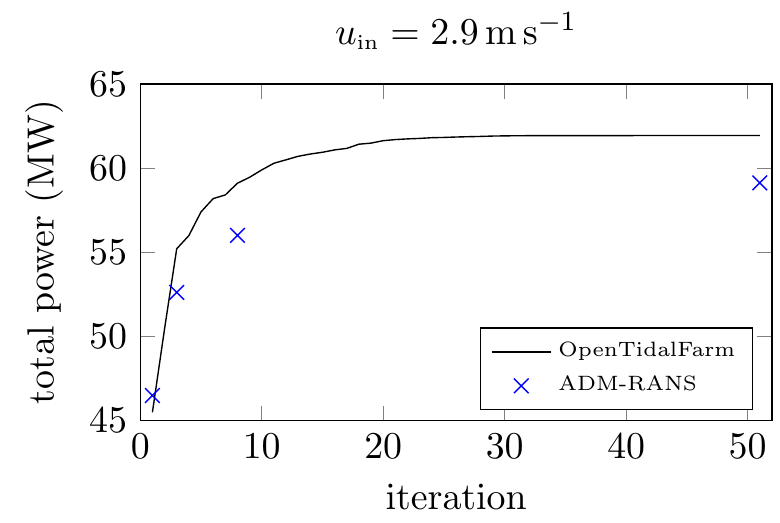}
    \caption{Total array power output at each OTF optimisation iteration. Iteration 1 represents the power output from the initial staggered layout. The crosses correspond to the power estimate obtained from a 3D Fluidity ADM-RANS simulation on the corresponding array layout.}
    \label{fig:OTF_vs_ADM_it}
  \end{figure}

  Fig.~\ref{fig:FL_32} illustrates the wakes formed behind the turbines in both the staggered and the optimised layouts along with a 2D slice through the 3D domain showing the optimised mesh. The total array extracted powers computed using Eq.~(\ref{eq:P3d}) are compared against the values obtained from the OTF simulations and the results are presented in Table~\ref{tab:power_comp} and Fig.~\ref{fig:OTF_vs_ADM_it}. The power predictions of the two models generally agree well with each other with a 8.5\% difference in the worst case. This difference is by no means insignificant, but it gives confidence that there is robustness in the improved array designs of OTF yielding increased power.\\
  
  \begin{table}[h!]
    \small
    \setlength{\tabcolsep}{8pt}
    \renewcommand{\arraystretch}{1.5}
    \begin{center}
      \begin{tabular}{|l c c c c|}\hline
         $u_\textnormal{in}$        & layout    & OTF power             & ADM-RANS power        & difference \\ \hline
         1.9 \si{\metre\per\second} & staggered & 12.98 \si{\mega\watt} & 14.18 \si{\mega\watt} & -8.5\%     \\ 
         1.9 \si{\metre\per\second} & optimised & 17.17 \si{\mega\watt} & 17.66 \si{\mega\watt} & -2.8\%     \\ 
         2.9 \si{\metre\per\second} & staggered & 45.48 \si{\mega\watt} & 46.48 \si{\mega\watt} & -2.2\%     \\ 
         2.9 \si{\metre\per\second} & optimised & 62.13 \si{\mega\watt} & 59.12 \si{\mega\watt} & +5.1\%     \\ \hline
      \end{tabular}
    \end{center}
    \caption{Steady flow simulation array power values comparing the variable viscosity OTF model against the 3D ADM-RANS model.}
    \label{tab:power_comp}
  \end{table}

  In order to further demonstrate the importance of using an appropriate viscosity value in the depth-averaged model, the OTF simulations were also run with a higher eddy viscosity value of $\nu=\SI{3}{\square\metre\per\second}$, as used in \cite{funke2014tidal}. The resulting initial staggered array power outputs were \SI{16.90}{\mega\watt} and \SI{57.87}{\mega\watt} for $u_\textnormal{in}=\SI{1.9}{\metre\per\second}$ and $u_\textnormal{in}=\SI{2.9}{\metre\per\second}$, respectively. These values are 19.2\% ($u_\textnormal{in}=\SI{1.9}{\metre\per\second}$) and 24.5\% ($u_\textnormal{in}=\SI{2.9}{\metre\per\second}$) greater than the values predicted by the ADM-RANS model. The reason for this lies in the fact that the higher viscosity used in OTF leads to faster wake recoveries which in turn results in greater velocities at the turbine locations. Given that the power output scales with velocity cubed, small differences in velocity are magnified in the power output values and this has led to exaggerated array power output values in OTF. This highlights the importance of correctly calibrating viscosity in the depth averaged model as small differences in $\nu$ will result in significant errors in the predicted array power outputs.

  \subsubsection{Power per turbine}
\begin{figure}[h!]
    \centering
    \begin{subfigure}[h!]{0.48\textwidth}
      \centering
      \includegraphics[width=\textwidth]{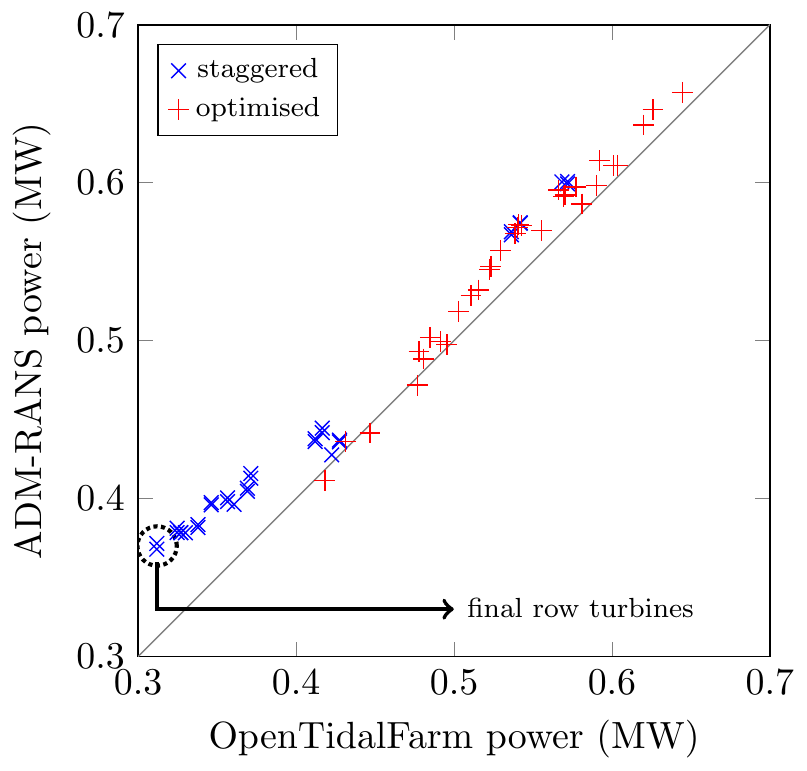}
      \caption{$u_\textnormal{in}=\SI{1.9}{\metre\per\second}$}
      \label{fig:OTF_vs_ADM_cor_19}
    \end{subfigure} \hfill
    \begin{subfigure}[h!]{0.48\textwidth}
      \centering
      \includegraphics[width=\textwidth]{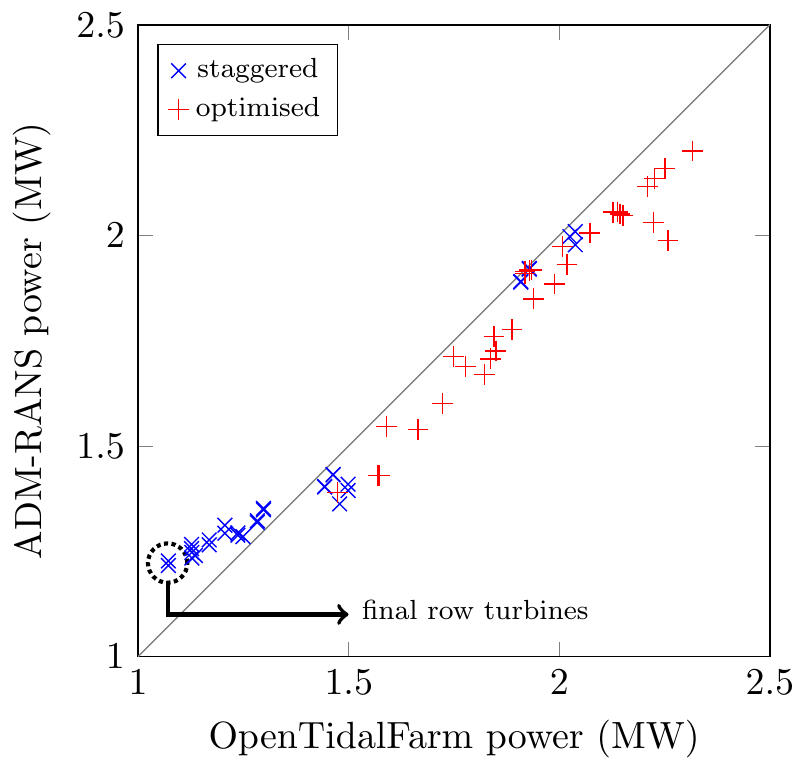}
      \caption{$u_\textnormal{in}=\SI{2.9}{\metre\per\second}$}
      \label{fig:OTF_vs_ADM_cor_29}
    \end{subfigure}
    \caption{OTF vs. Fluidity ADM-RANS power per turbine comparison plot for the staggered and optimised layouts. The diagonal line indicates a perfect match between the two models.}
    \label{fig:OTF_vs_ADM_cor}
  \end{figure}

  In order to examine the discrepancies between the two models, a closer examination of the extracted power values is presented in this section. A comparison between the power output per turbine predicted by the two models for both upstream velocity values is presented in Fig.~\ref{fig:OTF_vs_ADM_cor}. In this plot the power per turbine predicted by the 3D ADM-RANS model, Eq.~(\ref{eq:P3d}), is plotted against the value predicted by the OTF simulations, Eq.~(\ref{eq:P2d}). The results for both the initial staggered layouts and the final optimised layouts are shown. The diagonal line indicates a perfect match between the two models and therefore the results are judged based on how much they deviate away from this line.\\

  The OTF model has predominantly underpredicted the power values for the staggered layout with maximum differences of 20\% for the $u_\textnormal{in}=\SI{1.9}{\metre\per\second}$ case and 15\% for the $u_\textnormal{in}=\SI{2.9}{\metre\per\second}$ case. In fact, the differences between OTF and ADM-RANS are largest for the turbines with the lowest power outputs since these turbines are the ones located in the final row of the staggered layout. The reason for this lies in the fact that for the first row since there are no turbines upstream, inaccuracies in the wake profile predictions do not affect the power predictions. However, further downstream in the later rows, the discrepancies between the two models grow due to the limitations associated with the wake profile predictions of OTF. These inaccuracies are amplified and therefore the worst agreements can be observed in the final row of the staggered layout. This is not surprising given that the velocity values at these locations are heavily dependent on the values upstream and therefore small discrepancies upstream would be amplified at these locations.\\

  The results for the optimised layout are more balanced and the maximum differences are 6\% for the $u_\textnormal{in}=\SI{1.9}{\metre\per\second}$ case and 12\% for the $u_\textnormal{in}=\SI{2.9}{\metre\per\second}$ case. In these layouts, the turbines are not positioned in the wake of any upstream turbine and therefore inaccuracies in wake profile predictions of OTF are not magnified as much. Overall, given the assumptions used in this study and the different approaches used to model the turbines, a good agreement between the power values predicted by the two models can be observed. This is also reflected in the relatively small differences observed in the predicted total array power outputs, Table~\ref{tab:power_comp}.\\

  \subsection{Unsteady flow}
  Having shown good qualitative agreement between the wake profiles and the power production predicted by the two models for constant inlet velocities, in this section the analysis is extended to examine an unsteady scenario with a time dependent inlet velocity. Once again, the same ideal channel described in Fig.~\ref{fig:domain_32} is also used here. However, here the inlet velocity is modified to follow a sinusoidal profile such that

  \begin{equation}
    u_\textnormal{in}(t) = u_\textnormal{max} \sin( 2\pi (t/\tau) ) \, ,
  \end{equation}

  \noindent where $\tau=\SI{12}{\hour}$ and $u_\textnormal{max}=\SI{3.9}{\metre\per\second}$. A half-cycle is considered here and therefore flow reversal is not modelled. Moreover, the boundary conditions, apart from at the inlet, are identical to those described in section \ref{sec:nu_sens_bc}.

    \subsubsection{OpenTidalFarm -- depth-averaged}
    Previously, in the steady flow OTF scenarios considered, the array power output was maximised for the particular inlet velocity value. However, for unsteady simulations, OTF optimises the turbine positions in order to maximise the array power integral over time. Hence, the optimised positions reflect the array layout that yields the maximum power integral for the duration of the \SI{6}{\hour} half-cycle. Furthermore, in the OTF simulations presented here, the eddy viscosity coefficient is also time dependent and is set to vary with the inlet velocity according to the relationship derived in section \ref{sec:nu_sens}, Eq.~(\ref{eq:nu_best}). This will further examine the benefits of varying the viscosity in-line with the inlet velocity, as opposed to using a constant spurious viscosity value throughout.\\

\begin{figure}[h!]
    \centering
    \begin{subfigure}[h!]{0.45\textwidth}
      \centering
      \includegraphics[width=\textwidth]{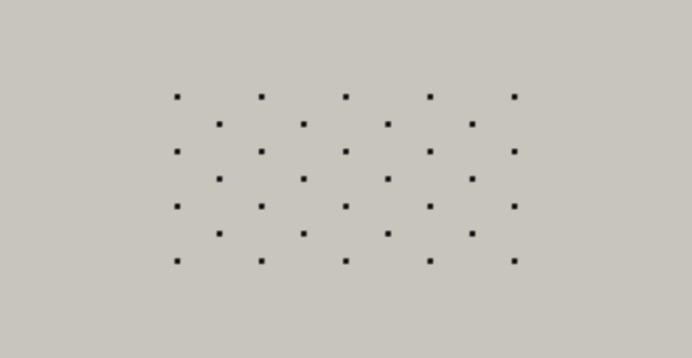}
      \caption{staggered turbine positions}
    \end{subfigure} \hfill
    \begin{subfigure}[h!]{0.45\textwidth}
      \centering
      \includegraphics[width=\textwidth]{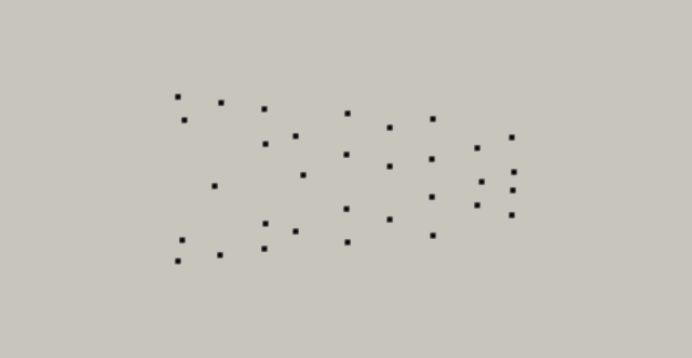}
      \caption{optimised turbine positions}
    \end{subfigure}
    \caption{OTF results showing the initial staggered layout on the left and the optimised layout on the right for the time-dependent inflow velocity case.}
    \label{fig:OTF_32_unst}
  \end{figure}

    Fig.~\ref{fig:OTF_32_unst} displays the optimisation results with the initial staggered turbine layout on the left and the optimised turbine layout on the right. As in section \ref{sec:channel_steady}, a minimum spacing of $2D$ between adjacent turbines was enforced in the OTF optimisation. As might be expected in this scenario without flow reversal, the optimised layout is very similar to the two optimised layouts observed for the steady flow cases considered, Fig.~\ref{fig:OTF_32_pos_19}--\ref{fig:OTF_32_pos_29}. Once again, OTF has avoided placing turbines downstream of each other as far as possible. Here, OTF predicts a power integral of \SI{281}{\mega\watt\hour} for the initial staggered layout which increases to \SI{386}{\mega\watt\hour} with the optimised layout. This is a substantial increase of 37\%.

    \subsubsection{Fluidity -- 3D ADM-RANS}
\begin{figure}[h!]
    \centering
    \includegraphics[scale=0.8]{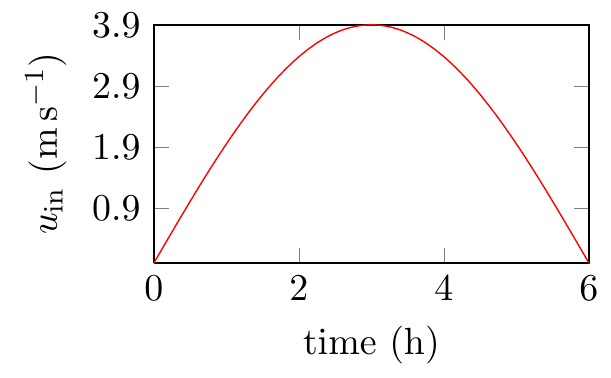}
    \includegraphics[scale=0.8]{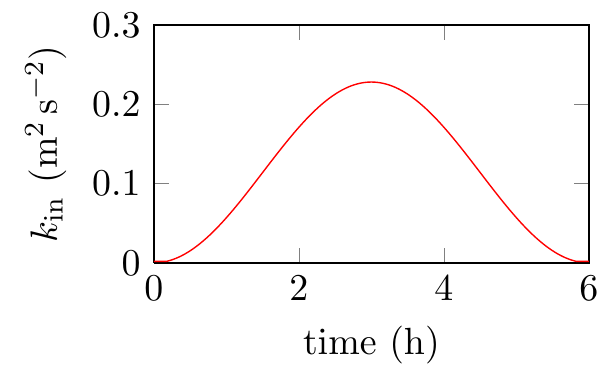}
    \includegraphics[scale=0.8]{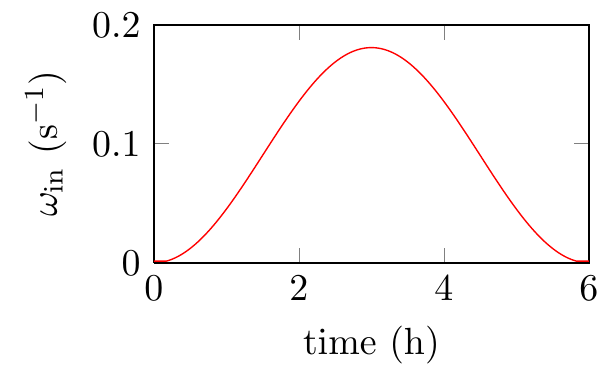}
    \includegraphics[scale=0.8]{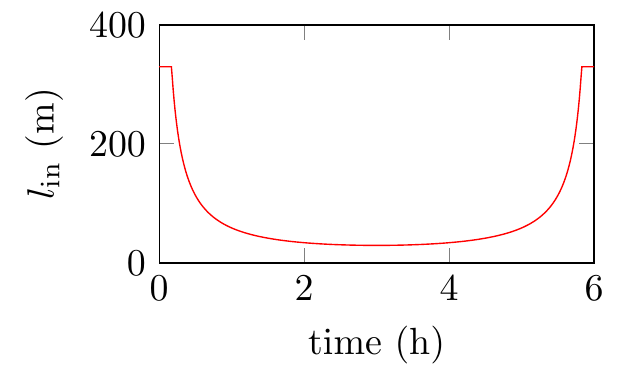}
    \caption{Variation of the inlet boundary conditions with time for the unsteady flow ADM-RANS simulations during the \SI{6}{\hour} half-cycle.}
    \label{fig:inlet}
  \end{figure}

    The 3D ADM-RANS model was again used to assess the OTF power predictions. In the ADM-RANS model, the turbulence properties (i.e. $k$ and $\omega$) were also assumed to be time dependent and follow scenario (c), described in section \ref{sec:nu_sens}, where $I$ and $\nu_T$ are unaffected by changes to the inlet velocity. The variation of inlet velocity, $k$ and $\omega$ with time is shown in Fig.~\ref{fig:inlet}. One of the issues with scenario (c) is that the turbulence length scale becomes very large for small values of $k$ and $\omega$. Therefore, minimum values of $k_\textnormal{min}=\SI{1.80e-3}{\square\metre\per\square\second}$ and $\omega_\textnormal{min}=\SI{1.43e-3}{\per\second}$ have been specified to ensure that the turbulence length scale is capped at $l_\textnormal{max}=\SI{330}{\metre}$, which is equivalent to 10 times the depth of the channel. This value is also less than half the channel width and it is not unreasonable to assume the existence of horizontal eddies with this length scale within the flow. The variation of $l$ at the inlet over time is also shown in Fig.~\ref{fig:inlet}.\\

\begin{figure}[h!]
    \centering
    \begin{subfigure}[h!]{0.45\textwidth}
      \centering
      \includegraphics[width=\textwidth]{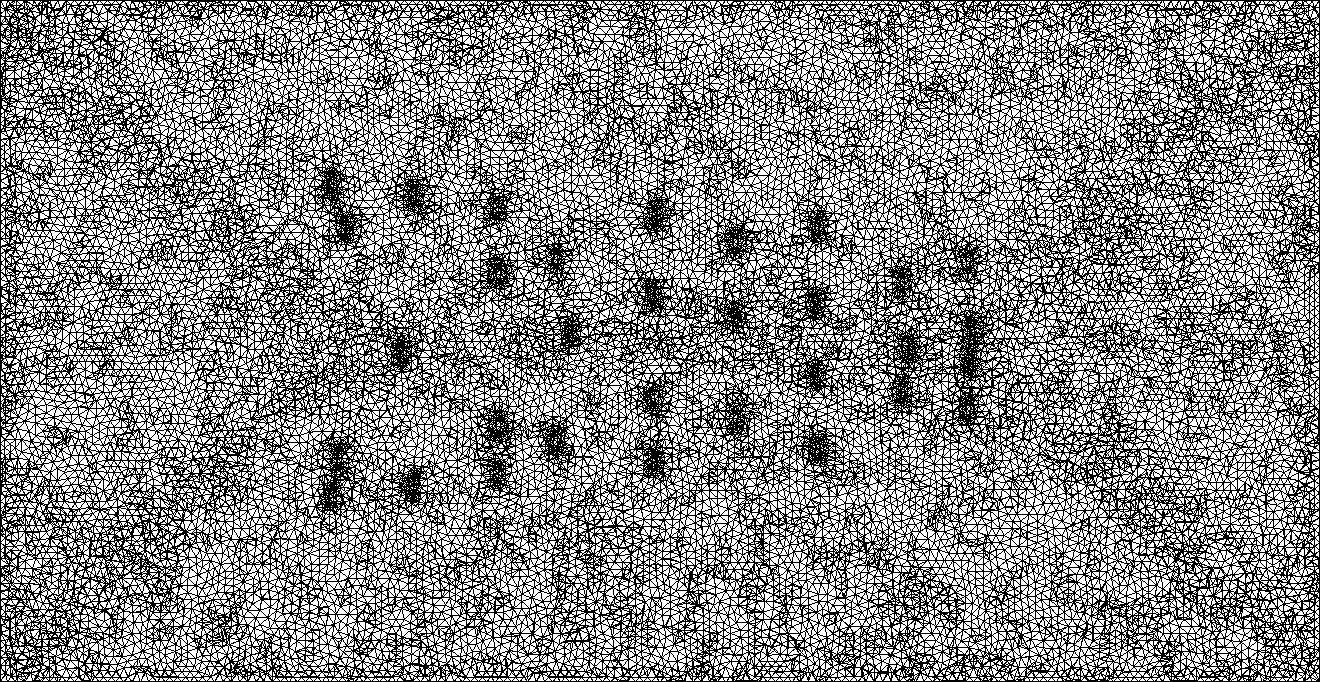}
      \caption{$t=\SI{0}{\hour}$ ($u_\textnormal{in}=\SI{0}{\metre\per\second}$)\\No. of elements: \SI{3.99e5}{}}
    \end{subfigure} \hfill
    \begin{subfigure}[h!]{0.45\textwidth}
      \centering
      \includegraphics[width=\textwidth]{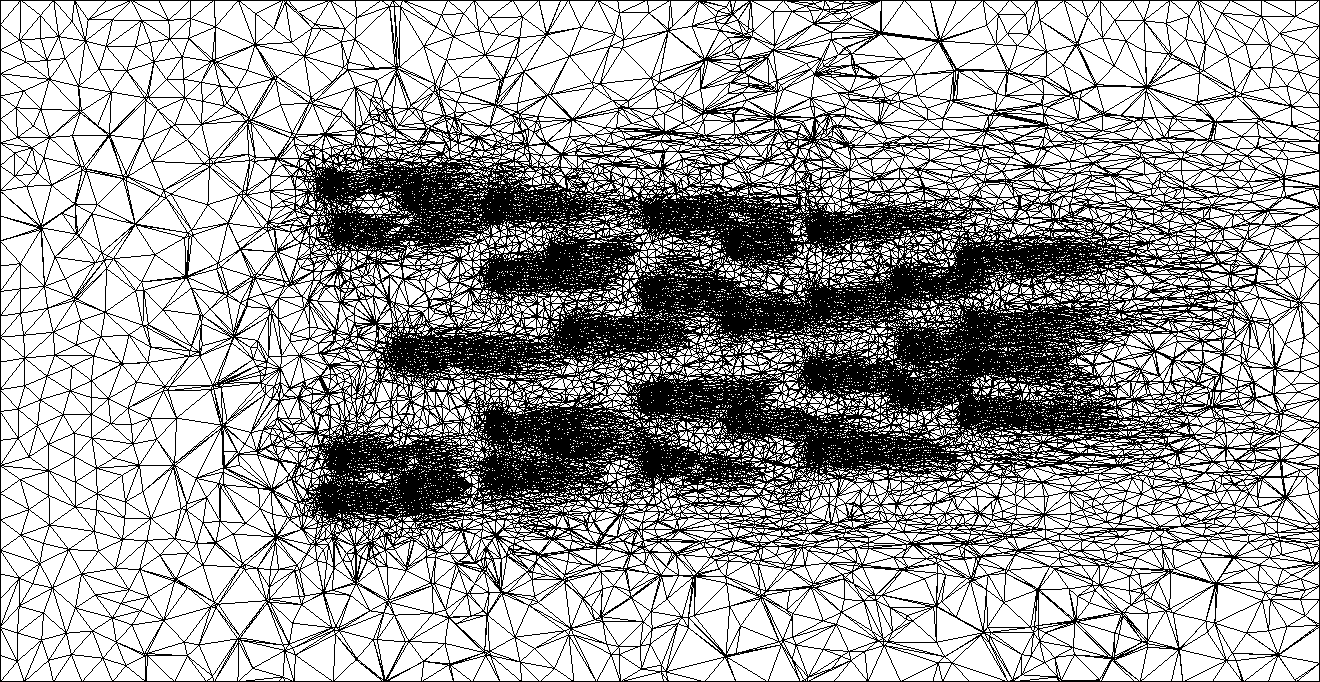}
      \caption{$t=\SI{0.9}{\hour}$ ($u_\textnormal{in}=\SI{1.77}{\metre\per\second}$)\\No. of elements: \SI{4.67e5}{}}
    \end{subfigure}
    \par\bigskip 
    \begin{subfigure}[h!]{0.45\textwidth}
      \centering
      \includegraphics[width=\textwidth]{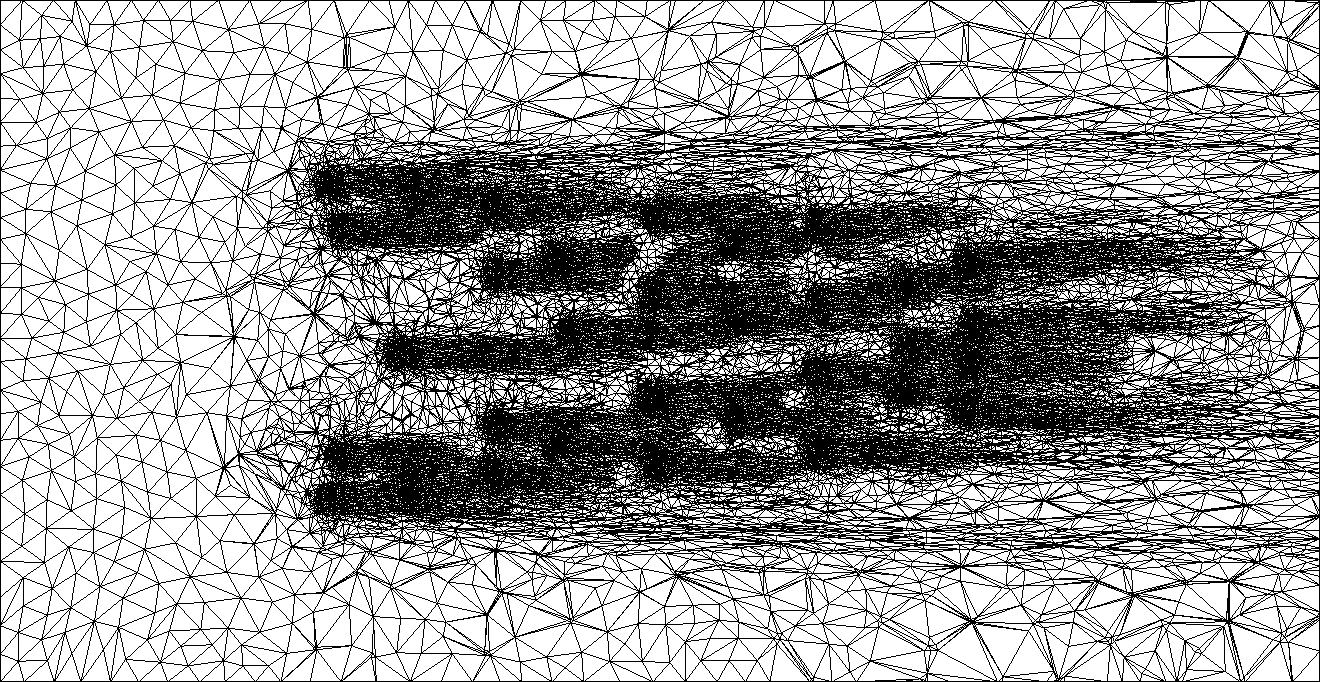}
      \caption{$t=\SI{1.8}{\hour}$ ($u_\textnormal{in}=\SI{3.16}{\metre\per\second}$)\\No. of elements: \SI{5.09e5}{}}
    \end{subfigure} \hfill
    \begin{subfigure}[h!]{0.45\textwidth}
      \centering
      \includegraphics[width=\textwidth]{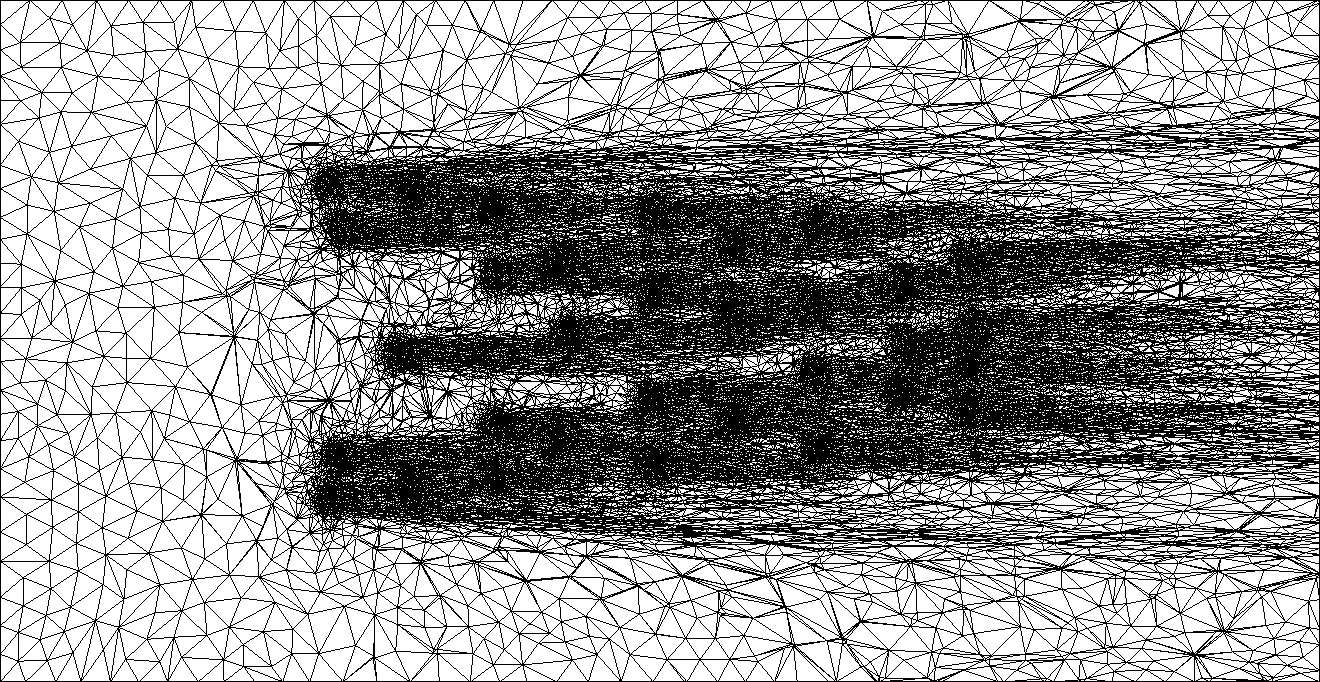}
      \caption{$t=\SI{3}{\hour}$ ($u_\textnormal{in}=\SI{3.9}{\metre\per\second}$)\\No. of elements: \SI{5.28e5}{}}
    \end{subfigure}
    \caption{2D slices at hub height through the 3D mesh at various times of the ADM-RANS unsteady flow simulation past the optimised layout.}
    \label{fig:mesh_opt_demo}
  \end{figure}

    Once again, mesh optimisation has been used with the same criteria described previously in section \ref{sec:nu_sens}. The mesh optimisation capabilities available in the ADM-RANS model help reduce the computational cost of these simulations. A high resolution initial mesh that anticipates the position and extent of the wakes is not required. Instead, the initial mesh simply needs to resolve the turbines (i.e. actuator discs) with sufficient resolution, saving time on the pre-processing. Thereafter, the model will, if and when necessary, increase the mesh resolution downstream of the turbines as the wakes start to develop. Thus, capturing the wake interactions with sufficient accuracy. In order to demonstrate this, 2D slices at hub height through the 3D domain showing the optimised mesh at various times of the unsteady simulation past the optimised layout are shown in Fig.~\ref{fig:mesh_opt_demo}. As demonstrated, the number of elements used increases with increasing $u_\textnormal{in}$ in order to ensure that the wakes downstream of the turbines are correctly captured.\\

\begin{figure}[h!]
    \centering
    \begin{subfigure}[h!]{0.45\textwidth}
      \centering
      \includegraphics[width=\textwidth]{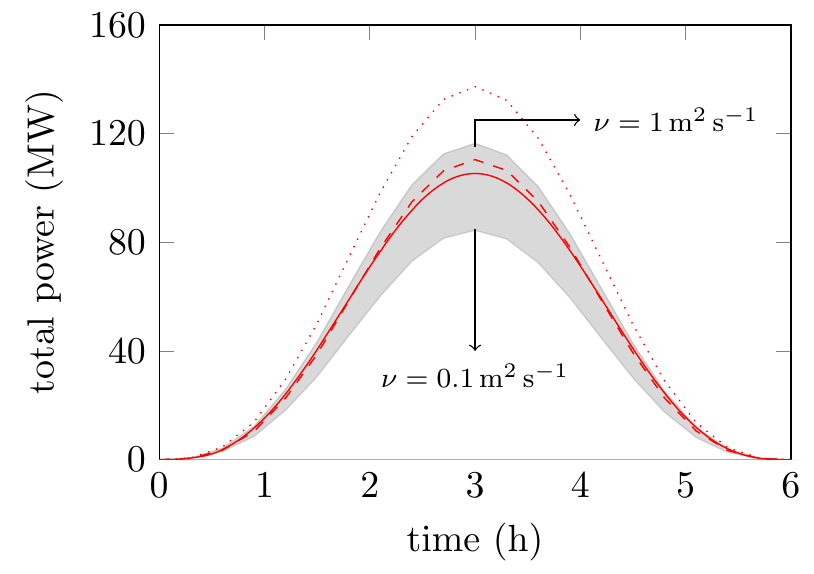}
      \caption{staggered layout}
      \label{fig:OTF_vs_ADM_unst_stg}
    \end{subfigure} \hfill
    \begin{subfigure}[h!]{0.45\textwidth}
      \centering
      \includegraphics[width=\textwidth]{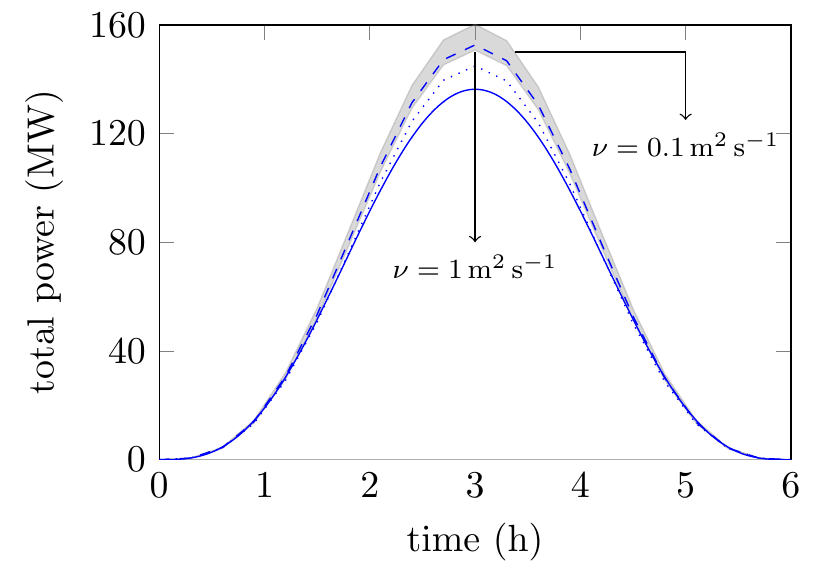}
      \caption{optimised layout}
      \label{fig:OTF_vs_ADM_unst_opt}
    \end{subfigure}
    \caption{OTF vs ADM-RANS unsteady flow. The solid lines show the ADM-RANS results (left (\protect\tikz \protect\draw[color=red] (0,10)--(0.5,10);) and right (\protect\tikz \protect\draw[color=blue] (0,10)--(0.5,10);) ) and the dashed lines show the OTF results with a variable viscosity (left (\protect\tikz \protect\draw[style=dashed,color=red] (0,10)--(0.5,10);) and right (\protect\tikz \protect\draw[style=dashed,color=blue] (0,10)--(0.5,10);) ). The dotted lines show OTF results with $\nu=\SI{3}{\square\metre\per\second}$ (left (\protect\tikz \protect\draw[style=dotted,color=red] (0,10)--(0.5,10);) and right (\protect\tikz \protect\draw[style=dotted,color=blue] (0,10)--(0.5,10);) ). The shaded areas represent the range of values obtained if constant eddy viscosity values of $\SI{0.1}{\square\metre\per\second}\leq\nu\leq\SI{1}{\square\metre\per\second}$ are used in OTF.}
    \label{fig:OTF_vs_ADM_unst}
  \end{figure}

  The 3D ADM-RANS model was used to simulate the time dependent flow past the initial staggered and the final optimised layouts. Fig.~\ref{fig:OTF_vs_ADM_unst} shows the total array power production values predicted by the ADM-RANS model, as well as OTF, for both layouts. OTF depth-averaged simulations were also run with $\nu=\SI{0.1}{\square\metre\per\second}$ and $\nu=\SI{1}{\square\metre\per\second}$ corresponding to the lower and upper bound eddy viscosity values considered in section \ref{sec:nu_sens}. Moreover, a high eddy viscosity value of $\nu=\SI{3}{\square\metre\per\second}$, as was used in \cite{funke2014tidal}, was also considered. The various power integral values obtained for the two layouts shown in Fig.~\ref{fig:OTF_32_unst} are presented in Table~\ref{tab:power_int}.\\
  
  \begin{table}[h!]
    \small
    \setlength{\tabcolsep}{8pt}
    \renewcommand{\arraystretch}{1.5}
    \begin{center}
      \begin{tabular}{|l l l|}\hline
         Layout                                         & staggered (\si{\mega\watt\hour}) & optimised (\si{\mega\watt\hour}) \\ \hline
         ADM-RANS                                       & 278.0                            & 357.0                            \\
         OTF (variable $\nu$)                           & 281.5 (+ 1.3\%)                  & 386.3 (+ 8.2\%)                  \\
         OTF ($\nu=\SI{0.1}{\square\metre\per\second}$) & 216.7 (-22.0\%)                  & 405.3 (+13.5\%)                  \\
         OTF ($\nu=\SI{1}{\square\metre\per\second}$)   & 302.1 (+ 8.7\%)                  & 379.9 (+ 6.4\%)                  \\
         OTF ($\nu=\SI{3}{\square\metre\per\second}$)   & 353.8 (+27.3\%)                  & 366.8 (+ 2.7\%)                  \\ \hline
      \end{tabular}
    \end{center}
    \caption{Unsteady flow simulation power integral values. The percentage difference of the OTF data relative to the ADM-RANS data is shown inside the brackets.}
    \label{tab:power_int}
  \end{table}

  As demonstrated earlier in section~\ref{sec:nu_sens}, increasing $\nu$ leads to shorter wakes in OTF. In the staggered case, apart from the first two rows, the turbines are positioned in the wakes of upstream turbines. Hence, shorter wakes leads to higher velocities at downstream turbine locations and this results in greater array power extraction values. Therefore, the OTF run with constant $\nu=\SI{3}{\square\metre\per\second}$ predicts the highest array power integral and overestimates the ADM-RANS value by more than 27\% (Table~\ref{tab:power_int}). On the other hand, the OTF run with constant $\nu=\SI{0.1}{\square\metre\per\second}$ predicts the lowest array power integral and underestimates the ADM-RANS value by 22\%. The closest match using the constant eddy viscosity OTF model is achieved with $\nu=\SI{1}{\square\metre\per\second}$ which overpredicts the power integral by almost 9\%. The best agreement between the depth-averaged OTF and the 3D ADM-RANS results is achieved using the variable eddy viscosity OTF simulation, Fig.~\ref{fig:OTF_vs_ADM_unst_stg}, which overpredicts the power integral by only 1\%.\\

  However, in the optimised layout the trend is reversed and the highest OTF power integral values are produced with constant $\nu=\SI{0.1}{\square\metre\per\second}$ while the OTF constant $\nu=\SI{3}{\square\metre\per\second}$ run leads to the lowest power integral value (Table~\ref{tab:power_int}). The reason for this lies in the fact that in the optimised layout the turbines are predominantly positioned in the bypass flow region of upstream turbines. Once again, increasing $\nu$ leads to shorter wakes, but this also leads to lower bypass flow velocities. Therefore, in the optimised layout, the velocities at the turbine locations decrease as $\nu$ increases. Hence, increasing $\nu$ leads to lower power integral values in OTF, Fig.~\ref{fig:OTF_vs_ADM_unst_opt}.\\

  Furthermore, in the optimised layout, all OTF simulations overpredict the ADM-RANS data, but the differences are smaller than those observed for the staggered layout. In fact, the $\nu=\SI{0.1}{\square\metre\per\second}$ OTF value and $\nu=\SI{3}{\square\metre\per\second}$ OTF value are only 10\% different relative to each other. Given that in this layout the turbines are not positioned directly behind one another, the knock on effect observed in the staggered layout does not apply here and the differences between the various runs are relatively smaller. These differences are primarily due to the 2D representation of truly 3D wake profiles. The match between the ADM-RANS model and the variable eddy viscosity OTF model can be improved by investigating the relationship between $\nu$ and $u_\textnormal{in}$ in greater detail. In this study, a simple linear relationship has been suggested, however this should be further examined in order to improve the agreement between the two models. Nonetheless, a good qualitative agreement between the 3D ADM-RANS model and the depth-averaged OTF model has been observed for both the steady and the unsteady flow cases considered in this study. This gives confidence in the value of the optimised layouts obtained via OTF, although higher fidelity 3D models may still be required to accurately estimate the array power outputs and to produce truly optimal array layouts.\\

\begin{figure}[h!]
    \centering
    \includegraphics[width=0.6\textwidth]{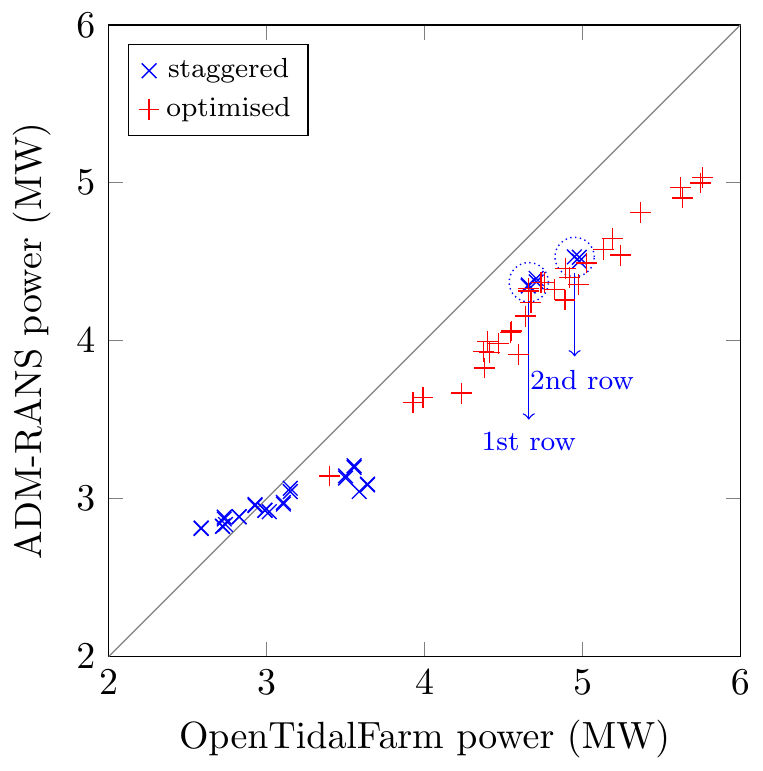}
    \caption{Variable $\nu$ OTF vs ADM-RANS power per turbine comparison plot at $t=\SI{3}{\hour}$ and $u_\textnormal{in}=\SI{3.9}{\metre\per\second}$.}
    \label{fig:OTF_vs_ADM_3.9_vary}
  \end{figure}

  In order to better understand the differences between the 3D ADM-RANS and the variable eddy viscosity OTF power output values, the power per turbine values at the maximum velocity point (i.e. $u_\textnormal{in}=\SI{3.9}{\metre\per\second}$ at $t=\SI{3}{\hour}$) are compared against each other and are presented in Fig.~\ref{fig:OTF_vs_ADM_3.9_vary}. For the staggered layout, there is a significant difference between the power outputs from the turbines in the first two rows compared to all other turbines. The first row benefits from a high velocity due to the undisturbed inflow. The second row is placed in the bypass flow region of the first row and therefore experiences an even higher velocity which results in even higher power values. All subsequent turbines are placed in the wake of upstream turbines and therefore experience much lower velocities. For the optimised layout, all the turbines lie along a line which is offset from the perfect diagonal. This suggests that better scaling can improve the match between OTF and the 3D ADM-RANS model. The relationship between $\nu$ and $u_\textnormal{in}$ derived in section~\ref{sec:nu_sens} was based on a simple linear fit. In order to come up with a more comprehensive relationship, the range of velocities considered needs to be extended and a higher order best-fit is required.\\

\begin{figure}[h!]
    \centering
    \begin{subfigure}[h!]{0.6\textwidth}
      \centering
      \includegraphics[width=\textwidth]{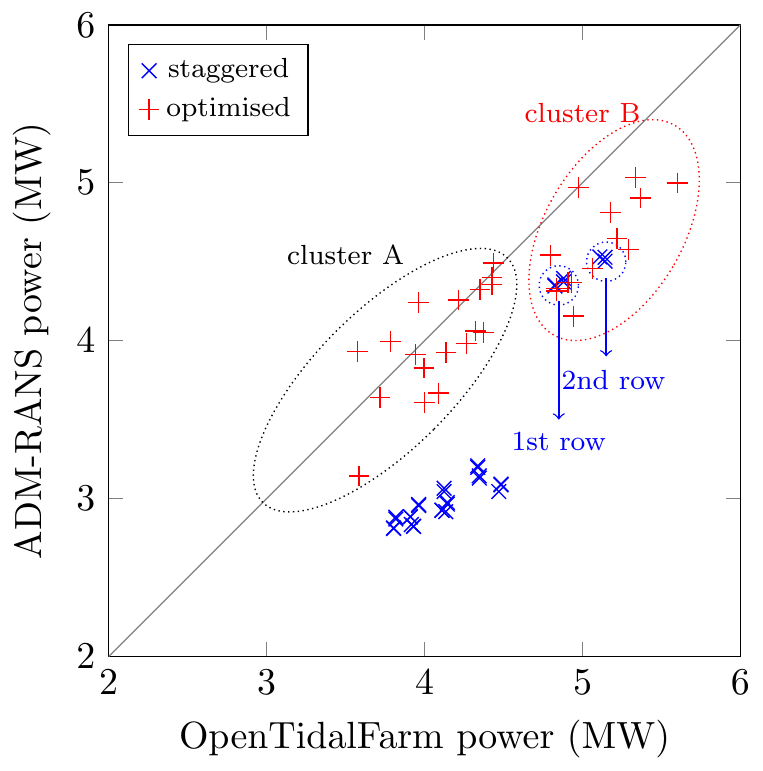} 
      \caption{Power per turbine}
      \label{fig:OTF_vs_ADM_3.9_cor}
    \end{subfigure} \hfill
    \begin{subfigure}[h!]{0.38\textwidth}
      \centering
      \includegraphics[width=\textwidth]{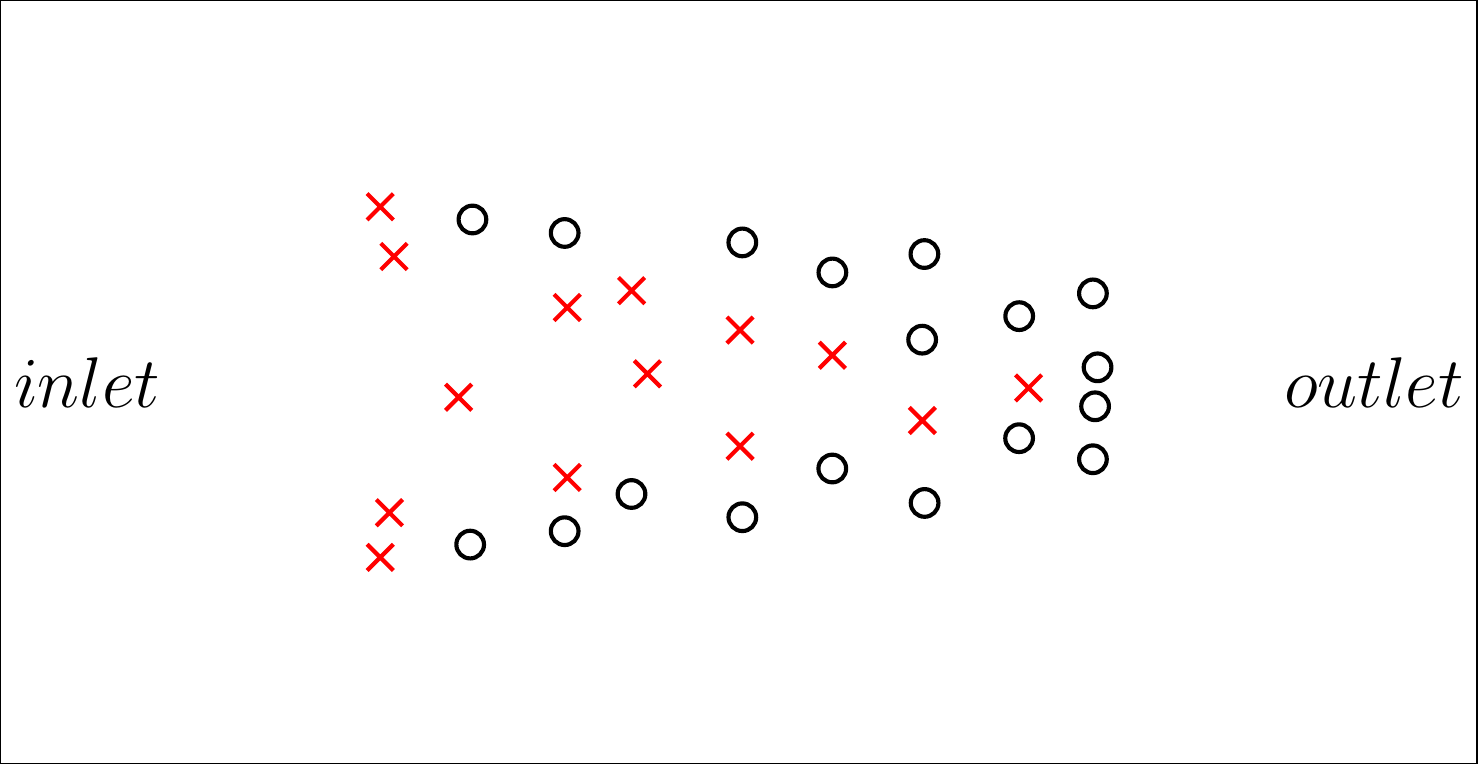}
      \caption{Optimised layout with cluster A turbines shown in circles (\protect\tikz \protect\draw[thick] circle (1.5pt);) and cluster B turbines shown in crosses (\protect\tikz \protect\draw[color=red,mark=x] plot(10cm,10cm);).}
      \label{fig:OTF_vs_ADM_3.9_pos}
    \end{subfigure}
    \caption{Constant $\nu=\SI{3}{\square\metre\per\second}$ OTF vs. ADM-RANS power per turbine comparison plot.}
    \label{fig:OTF_vs_ADM_3.9}
  \end{figure}

\begin{figure}[h!]
    \centering
    \includegraphics[width=0.6\textwidth]{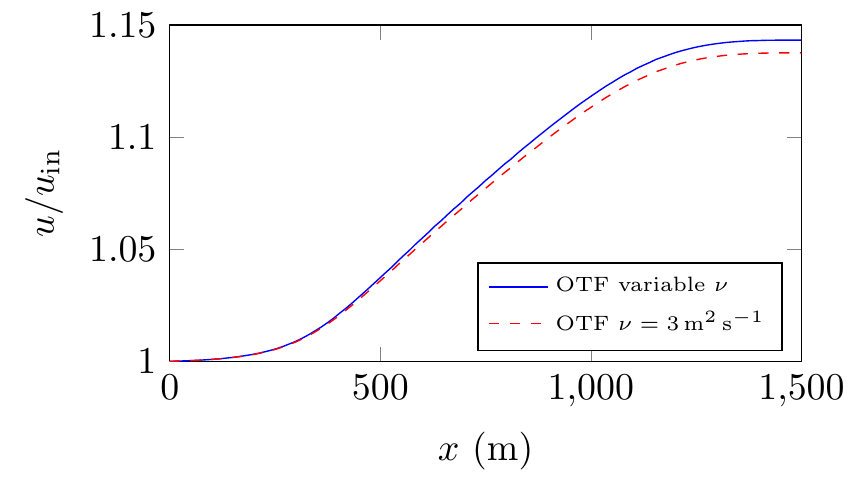}
    \caption{Normalised velocity along the channel side walls for two different OTF simulations at $t=\SI{3}{\hour}$ and $u_\textnormal{in}=\SI{3.9}{\metre\per\second}$. The channel inlet is at $x=\SI{0}{\metre}$ and the outlet is at $x=\SI{1500}{\metre}$.}
    \label{fig:OTF_sidewall}
  \end{figure}

  According to Fig.~\ref{fig:OTF_vs_ADM_unst_opt}, the closest match to the ADM-RANS model for the optimised layout is provided by the constant $\nu=\SI{3}{\square\metre\per\second}$ OTF model. Fig.~\ref{fig:OTF_vs_ADM_3.9_cor} presents the power per turbine values at the maximum velocity point predicted by this model against the ADM-RANS power values. The staggered data is very similar to that presented in Fig.~\ref{fig:OTF_vs_ADM_3.9_vary} with the first two rows of turbines producing significantly more power than the rest of the array. For the optimised layout, the results can be divided into two clusters. The power values for the turbines in cluster A, identified in Fig.~\ref{fig:OTF_vs_ADM_3.9_vary}, are well predicted by the constant $\nu=\SI{3}{\square\metre\per\second}$ OTF model, but those in cluster B are consistently overpredicted. Fig.~\ref{fig:OTF_vs_ADM_3.9_pos} shows the positions of the turbines in the optimised layout with the two clusters identified. The turbines in cluster A help direct the flow down the middle of the channel, which is where the cluster B turbines are positioned. Hence, the cluster B turbines experience a higher inflow velocity and produce more power than the turbines in cluster A. Moreover, note that most of the turbines in cluster A are positioned close to the channel side walls in the channel bypass flow region. The velocity in the bypass flow region decreases slightly for the constant $\nu=\SI{3}{\square\metre\per\second}$ OTF model compared against the variable $\nu$ OTF simulation, Fig~\ref{fig:OTF_sidewall}. The difference in velocity between the two OTF runs is only around 1\%, but given that power scales as velocity cubed, Eq.~(\ref{eq:P2d}), the smallest variations in velocity will have a significant effect on the turbine power values. Consequently, this decrease in bypass flow velocity in the constant $\nu=\SI{3}{\square\metre\per\second}$ OTF run helps decrease the power values for the turbines in cluster A and improves their match with the ADM-RANS data. Therefore, the constant $\nu=\SI{3}{\square\metre\per\second}$ OTF model is able to provide a closer match to the array power values than the variable $\nu$ OTF model. However, note that this only true for the optimised layout and that the staggered array power values are grossly exaggerated if a spurious viscosity value is used in OTF.\\

  Furthermore, although the $\nu=\SI{1}{\square\metre\per\second}$ and $\nu=\SI{3}{\square\metre\per\second}$ OTF runs provide a better match than the variable $\nu$ OTF run in the optimised layout, one should bear in mind that this optimised layout was actually achieved via the use of the variable viscosity relationship, Eq.~(\ref{eq:nu_best}), in OTF. For example, with $\nu=\SI{3}{\square\metre\per\second}$ in OTF, there is very little wake interaction in the initial staggered layout and the turbine positions would be hardly altered by OTF. Hence, the optimised layout obtained by using a spurious viscosity value in OTF is unlikely to be very different from the initial staggered layout. Note how the staggered and optimised power integral values for the $\nu=\SI{3}{\square\metre\per\second}$ OTF runs are only 3.6\% different from each other. Therefore, an optimised layout that leads to a 28.4\% increase in power integral (according to the ADM-RANS model) would not have been obtained without the use of the variable viscosity OTF model. Hence, it has been demonstrated that employing a suitable calibrated viscosity value consolidates and improves the array optimisation of OTF and helps obtain more realistic optimised layouts.\\

  One of the main limitations of the current study is that global blockage effects have effectively been neglected in the channel flow cases considered: the inlet velocity is assumed to be independent of the flow blockage inside the channel. Hence, the inlet velocity is unaffected by the turbine layouts. However, in a realistic environment, this is not the case and the upstream velocity is reduced as more and more turbines are added and this will depend on the turbine layout. Therefore, the next step would be to apply a more realistic boundary condition (e.g. fixed head difference) at the inlet. Given that a Dirichlet velocity boundary condition was used in both the 3D ADM-RANS and the depth-averaged OTF simulations, the comparisons presented here are still of great value. Switching to a fixed pressure/head difference boundary condition will alter the power production values and change the optimised layouts, but the qualitative agreement established between 3D ADM-RANS and the depth-averaged OTF models should still hold true, as long as the boundary conditions used in the two models are consistent.\\

\section{Conclusions}
In order to assess the suitability of using depth-averaged models to assess the flow past arrays of tidal turbines, a comprehensive comparison between the wake profiles and array power outputs predicted by OTF and the 3D Fluidity ADM-RANS model has been presented. A viscosity sensitivity analysis has been carried out to outline the limitations associated with the use of a constant viscosity everywhere in the domain in the depth-averaged model. This has highlighted the need to adjust the viscosity value used in the depth-averaged model in-line with the upstream velocity value. Based on this, a simple relationship between viscosity and the upstream velocity value is suggested for the channel flow case.\\

Furthermore, this relationship was used to compute the appropriate viscosity value to be used in OTF for the flow past an array of 32 tidal turbines in an ideal channel. Both steady and unsteady flow cases have been examined. The OTF package was then used to optimise the positions of the 32 turbines to maximise the power output from the array. Several array designs from the OTF 32 tidal turbine array simulations have been replicated using the Fluidity 3D ADM-RANS model and the total power outputs predicted by the two models have been compared against each other. This allowed for an assessment into the accuracy of the adjoint based optimisation used in OTF for the first time. The results have outlined a good qualitative agreement between the two models, highlighting the importance of the correction to the power calculations employed in OTF as well as the viscosity relationship suggested. This agreement only holds true if appropriate viscosity values are chosen, highlighting the importance of careful viscosity calibration.\\

In conclusion, it is not always feasible to incorporate accurate and appropriate turbulence models into large scale tidal flow simulations. Herein, it has been demonstrated that by using a simple linear relationship for viscosity that scales with upstream velocity, as opposed to using a spuriously high constant value, the disadvantages of not modelling for turbulence directly can be minimised. However, bathymetry effects have been neglected in this study and these will undoubtedly play a crucial role in the ambient turbulence values and consequently the rate of wake re-energisation. Therefore, in order to come up with a comprehensive viscosity definition to be used in large scale simulations, the analysis presented here should be extended to investigate the relationship between viscosity and bathymetry as well. A viscosity coefficient that scales with both upstream velocity and bathymetry will help limit the disadvantages of not accounting for turbulence even further. This will lead to significant savings in the computational cost of the large scale simulations.

\section*{Acknowledgement}
This work was supported by a studentship for the first author which has been funded by an EPSRC Industrial CASE award in collaboration with MeyGen Ltd. The authors further acknowledge the support of the Imperial College London High Performance Computing Service.

\appendix

\section*{References}
  \bibliographystyle{elsarticle-num} 
  \bibliography{elsarticle-template-num}




\end{document}